\documentclass[mnsc]{informs3nothing}

\OneAndAHalfSpacedXI
\RequirePackage{
amsmath,
times,
graphicx,
amsfonts,
amscd,
amssymb,
algorithm,
algorithmic
}
\usepackage{natbib}
 \bibpunct[, ]{(}{)}{,}{a}{}{,}%
 \def\bibfont{\small}%
 \def\bibsep{\smallskipamount}%
 \def\bibhang{24pt}%
\TheoremsNumberedThrough     % Preferred (Theorem 1, Lemma 1, Theorem 2)
\ECRepeatTheorems

\EquationsNumberedThrough    % Default: (1), (2), ...
\usepackage{tabulary}

\newcommand{\paranth}[1]{\left(#1\right)}
\newcommand{\bracket}[1]{\left[#1\right]}
\newcommand{\curly}[1]{\left\{#1\right\}}

\numberwithin{equation}{section}
\theoremstyle{plain}
\newtheorem{thm}{Theorem}[section]

\newtheorem{lem}{Lemma}

\graphicspath{
{C:/Users/Steamed/Dropbox/RankBomb/PaperRankbomb/CT_model_results/}{C:/Users/Steamed/Dropbox/RankBomb/PaperRankbomb/Convergence_plot/convergence_user_by_user/4_30_ss_20000_perm_10_four_category/}{C:/Users/Steamed/Dropbox/RankBomb/PaperRankbomb/Convergence_plot/convergence_pairwise_data/2015_5_3/}{C:/Users/Steamed/Dropbox/RankBomb/PaperRankbomb/2015_3_8_fix_an_error/pathological_cases/}
{C:/Users/Steamed/Dropbox/RankBomb/PaperRankbomb/Figures/}
{C:/Dropbox/RankBomb/PaperRankbomb/CT_model_results/}{C:/Dropbox/RankBomb/PaperRankbomb/Convergence_plot/convergence_user_by_user/4_30_ss_20000_perm_10_four_category/}{C:/Dropbox/RankBomb/PaperRankbomb/Convergence_plot/convergence_pairwise_data/2015_5_3/}{C:/Dropbox/RankBomb/PaperRankbomb/2015_3_8_fix_an_error/pathological_cases/}
{C:/Dropbox/RankBomb/PaperRankbomb/Figures/}{D:/Dropbox/RankBomb/PaperRankbomb/Figures}
}

\begin{document} 

\RUNTITLE{Learning Preferences and User Engagement Using Choice and Time Data}
\TITLE{Learning Preferences and User Engagement Using Choice and Time Data}

\ARTICLEAUTHORS{
\AUTHOR{Zhengli Wang}
\AFF{Department of Mathematics, Massachusetts Institute of Technology, Cambridge MA 02139, \EMAIL{wzl@mit.edu}}
\AUTHOR{Tauhid Zaman}
\AFF{Sloan School of Management, Massachusetts Institute of Technology, Cambridge MA 02139, \EMAIL{zlisto@mit.edu}}
} % end of the block

\ABSTRACT{Choice decisions made by users of online applications can suffer from biases due to the users' level of engagement.
For instance, low engagement users may make random choices with no concern for the quality of items offered.
This biased choice data can corrupt estimates of user preferences for items.  However, 
one can correct for these  biases
if additional behavioral data is utilized.  To do this we construct
 a new \emph{choice engagement time} model which captures the impact of user 
engagement on choice decisions and response times associated with these choice decisions.
Response times are the behavioral data we choose because they are easily measured
by online applications and reveal information about user engagement.
To test our model we conduct online polls
with subject populations that have different levels of engagement and measure their choice decisions 
and response times. We have two main empirical findings.  First,
choice decisions and response times are correlated, with strong preferences having faster response times
than weak preferences.  Second,  low user engagement  is manifested through more random choice data
and faster response times.  Both of these phenomena are captured by our
choice engagement time model and we find that this model fits the data better than
traditional choice models.   
Our work has direct implications for online applications.  It lets these applications remove the bias
of low engagement users when estimating preferences for items.  It also allows for the segmentation of users according to their level of engagement, which can be useful for targeted advertising or marketing campaigns.  
}
%\HISTORY{This paper was first submitted on April 12, 1922 and has been with the authors for 83 years for 65 revisions.}

\maketitle
%%%%%%%%%%%%%%%%%%%%%%%%%%%%%%%%%%%%%%%%%%%%%%%%%%%%%%%%%%%%%%
\section{Introduction}
An important task in many applications is to determine the preferences of users for items.  Preference data is obtained by showing a user an offer set of items, and then letting the user choose the most preferred item.  This choice data is then used to learn the 
underlying preferences of users.  In many industries such as retailing, choice data is used to determine optimal sets of items to offer in order to maximize expected revenue.  In online dating applications such as Tinder or OkCupid, choice data is used to  find potential romantic partners for users.  Content sites such as YouTube and Netflix,  and social network sites such as Facebook, use choice data to select the best content to show users.  

For online applications, typically the user is shown choices (retail products, videos, news stories, romantic matches) and then clicks on a preferred option.  This is the choice data the application measures.  The issue is that this data may be biased by the engagement level of the users.  For instance,
a low engagement user may not even look at the items and choose randomly.  Or they may say ``yes'' to every item shown to them. 
This is the problem encountered in the dating application Tinder.  Many users, especially men, try to game the system by simply choosing ``yes'' for every person they are shown \citep{ref:tinder_swiperight}.  They are not strongly engaged with the decision making process and as a result are apathetic to the quality of the potential matches shown to them.  Rather, their broader strategy is to obtain as many matches as possible, for which saying yes to every match is the appropriate approach.  The problem with this behavior is that by liking every user shown as a potential match, the estimates of the quality of these users is biased upward.  This would result in lower users being given higher scores, which would result in them being shown to more often.  The effect of this would be to reduce the overall quality of the application since users would be shown lower quality matches.   If there was a way to detect and correct for this lack of engagement, then these types of users would not be able to bias the estimated quality of the potential matches. 

One feature that could help measure engagement is the response time associated with each choice decision.
This is the elapsed time between when the user sees the choices and when the user clicks on the preferred choice.
Response times are available for nearly every choice made in online applications.  
The ubiquity of response time data leads to several interesting questions.  
First, what is the relationship between choice and response times? 
Second, if such a relationship exists, how can it be used in order to better learn user preferences?  
And third, can choice and response time data provide behavioral insights about the user, such as their level of engagement?

The relationship between choice and response time  can be easily understood in the context of online applications.  For example, consider a dating application such as Tinder that shows a user a sequence of potential romantic matches.  Tinder is a rapid response type of application where the choice decision is based primarily on the photograph of the user.  Decisions are typically made in a few seconds without in-depth investigation of other information about the potential match, though this information is available by clicking on the photograph.  Typically, if the user sees someone that he strongly likes, then he would quickly select ``yes'' for this person.  Likewise, if the user sees someone he strongly dislikes, then we would quickly select ``no''.  However, if he is shown someone that he does not strongly like or dislike, then typically he will take more time to decide if he wants to select ``yes'' or ``no'' because it is more difficult to decide for such indeterminate cases.  Therefore, the response time  data would indicate that that there is a strong preference one way or another for the first two people, and no strong preference of the third person.  By formally modeling this relationship, response time  data can potentially be used to enhance basic choice models and learn preferences more accurately.  

There is also a natural relationship between choice, response time, and user engagement.  Consider again the example of Tinder.  
How can we distinguish unengaged users from  engaged users?  The choice data could help make this distinction, but it may not be sufficient.  For instance, while unengaged users may always select ``yes'' for every match, this type of behavior could also be due to a user that genuinely likes all matches shown.  However, there may be information in the response times that can reveal these users' lack of engagement.  If the actual image of the match does not affect the choice decision, then we would expect response times to be very similar for all matches shown.  Also, since there is essentially no decision process involved, the average response time would be faster than for normal engaged users.  If we could establish that such a phenomenon does occur and properly model it, then we would not only be able to correct for the bias of these unengaged users, we could also identify who they are, allowing for a useful segmentation of the user base.

%%%%%%%%%%%%%%%%%%%%%%%%%%%%%%%%%%%%%%%%%%%%%%%%%%%%%%%%%%%%%%%%%%%%
\subsection{Our Contributions}
In this work we present a model for the choice and response time data of individuals who are presented two choice options.  Our model is based on the relationship between choice, engagement, and time found in data from online experiments that we conduct.  The model we present is based on psychological models of decision processes, yet is  very simple and  tractable for estimation purposes.  In addition to allowing one to learn user preferences, our model is able to quantify the level of engagement of an individual user. We now discuss the main contributions of this work. 

%%%%%%%%%%%%%%%%%%%%%%%%%%%%%%%%%%%%%%%%%%%%%%%%%%%%%%%%%%%%%%%%%%%%%%%
\textbf{Empirical Analysis.}  We conduct online polls on two subject populations: students  at a college who take the
poll supervised by a researcher and Amazon Mechanical Turk (AMT) workers who take the polls from remote locations
with no supervision.  The polls allow us to obtain a complete set of pairwise comparisons of all poll items for all users.  Our analysis of this polling data reveals three important findings.
\begin{enumerate}
	\item   The measured choice fractions (defined as the minimum of the fraction of times an item is chosen in a given pair of items) are significantly closer to 0.5, which corresponds to random choice, for the AMT workers than for the students. 
	\item   The measured response times are significantly faster for the AMTs than the UGs. 
	\item   The measured correlation between choice fractions and response times is positive and significant for both AMT and
	student populations, but the correlation is substantially higher for the students.
\end{enumerate}
These findings suggest the following points.  First, it should be possible to jointly model choice and response time data given the observed correlations.  Second, there appears to be a relationship between choice, response time, and engagement.  In our data, the engagement is lower for the AMT workers because of their faster response times, more random choices, and lower correlation between choice fraction and response time.  This makes intuitive sense because AMT workers are online workers who may be less motivated or interested in the polling task than the students who were supervised by a researcher.

%%%%%%%%%%%%%%%%%%%%%%%%%%%%%%%%%%%%%%%%%%%%%%%%%%%%%%%%%%%%%%%%%%%%%%%
\textbf{Choice Engagement Time Model.}   
Motivated by our empirical findings, we develop a  Bayesian hierarchical choice engagement time model for choice and response time data which incorporates user engagement.  The model modifies the common multinomial-logit (MNL) choice model by including a user specific engagement parameter in addition to the normal item utility parameters.  The response times are modeled as the sum of a latency time and a decision time.  The latency time models the time to recognize the items shown and the decision time models the elapsed time for the decision process of the user once the items are recognized.  The latency times are independent of the item utilities, but the decision times depend upon the item utilities, user engagement, and a set of additional user parameters which weigh the relative importance of item utilities and user engagement in the decision time.  Our hierarchical modeling allows us to estimate user specific engagement and timing parameters.  The model parameters are estimated using standard Bayesian techniques.  We find that the choice engagement time model is a better fit to our observed choice and time data than traditional choice models.  In addition, we find that the estimated parameter values provide useful behavioral insights about the users' engagement
levels.

%%%%%%%%%%%%%%%%%%%%%%%%%%%%%%%%%%%%%%%%%%%%%%%%%%%%%%%%%%%%%%%%%%%%%%%%%%%%%%%%%%%%%%%%%%%%%

\subsection{Related Work}

Discrete choice models have been studied extensively in a wide variety of fields such as economics, transportation, psychology, marketing, and operations.  The majority of choice models consider only choice data.  Questions here include how to estimate such models and how to optimize item assortments with these models.  In the area of psychology the question of interest is in modeling the mental decision process to explain observed choice and response time data in laboratory experiments.

Early discrete choice models were based on random utility theory \citep{thurstone1927law}.  Each item $i$ was given a utility $u_i+\delta_i$ where $u_i$ is dependent upon features of the item and $\delta_i$ is a noise term used to model errors in the choice decision.  The noise term determines the distribution of the item utility and the resulting choice probabilities.  If the noise terms are normally distributed, the resulting model is known as the probit model.  If instead the noise term has an extreme value distribution, then it is known as the Bradley-Terry-Luce model \citep{bradley1952rank}.  Later this model also became known as the multinomial logit (MNL) model  \citep{mcfadden1973conditional} when item features were included. The MNL model has been the preferred model in transportation \citep{ben1985discrete} and marketing \citep{chandukala2008choice}.  In operations the MNL model has been very popular because its estimation and the corresponding assortment optimization problem are tractable \citep{talluri2004revenue}.  

The MNL model suffers from the phenomenon of Independence of Irrelevant Alternatives (IIA) 
\citep{debreu1960individual}, which states that the relative preference between
two choices will not change if additional alternatives are offered.  To overcome this limitation,
generalizations to the MNL model have been proposed such as the nested logit model \citep{domencich1975urban}, the general attraction model \citep{gallego2014general}, the Markov chain based choice model \citep{blanchet2016markov}, and distributions over permutations \citep{farias2013nonparametric}.   These more complex models are able to model a richer class of choice behavior, but their estimation and assortment optimization solutions are more complex compared to the simple MNL model.

The models discussed above only considered choice data because during the times when these models were first proposed, that was the only data available in most settings.  However, in parallel with the development of these pure choice models, psychologists developed choice models that also modeled the response time for the choice process.  These choice time models were based on the mental process underlying decisions for two item choices.  The earliest models were based on the sequential probability ratio test \citep{wald1948optimum}  where information was sequentially accumulated for the two items  and then a decision was made when a threshold was reached. Variations of this sequential sampling model include the accumulator model \citep{smith1988accumulator}, the recruitment model \citep{laberge1962recruitment}, and the runs model \citep{audley1965some}.  

Models were also developed which viewed the information accumulation process as a random walk  \citep{stone1960models,laming1968information,link1975sequential}.  The random walk modeled the relative information accumulation of each item, whereas previous models had focused on independent information accumulation processes for each item. The drift of the walk depends on the item utilities and there are decision thresholds above and below the starting point of the process.  An item is chosen when one of these thresholds is hit by the process, and the item chosen depends on whether the upper or lower threshold is hit first.  A continuous time version of these models known as the drift diffusion model was proposed by \cite{ratcliff1978theory}, where the information accumulation  process is modeled as a one dimensional Brownian motion.     The drift diffusion model has been successful at modeling phenomena seen in empirical data \citep{philiastides2010mechanistic, basten2010brain,ratcliff2002diffusion,ratcliff2000diffusion,ratcliff1997counter,mckeeff2007timing,leite2010modeling,domenech2010decision}.  However, a challenge posed by the drift diffusion model is the lack of a simple closed form expression for the likelihood function of the response times.  To overcome this, approximations have been developed to allow for easier model estimation \citep{navarro2009fast}.
Recent work has extended the  drift diffusion model to multi-item choice decisions \citep{krajbich2011multialternative}.  

%The modeling of the neurological process underling choice and response times has begun to impact fields outside of psychology, such as economics. It is argued in  \cite{krajbich2014benefits} that such neurological based models of choice can provide a better understanding of the true way people make choices, which can result in economic theories that more closely model real human behavior. Such an approach has already provided interesting insights and challenges to accepted economic models \citep{krajbich2015rethinking}. However, to the best our our knowledge, there has not been any substantial work in showing the value of such models for operational problems which rely upon choice models.
%

%%%%%%%%%%%%%%%%%%%%%%%%%%%%%%%%%%%%%%%%%%%%%%%%%%%%%%%%%%%%%%%%%%%%%%%%%%%%%%%%%%%%%%%%%%%%%
\subsection{Outline}
The remainder of our paper is structured as follows.  In Section \ref{sec:data} we describe our experiment and present exploratory data analysis.  We review previous choice time models in Section \ref{sec:previous_models}.  
Our empirical observations motivate our choice engagement time model which we present in 
Section \ref{sec:cet_model}.  Section \ref{sec:estimation} details our full hierarchical Bayesian model specification and estimation procedure.     In Section \ref{sec:results} we present model estimation results on our empirical choice and time data, along with comparisons with other models.  We summarize our main findings and conclude in Section \ref{sec:conclusion}.

%%%%%%%%%%%%%%%%%%%%%%%%%%%%%%%%%%%%%%%%%%%%%%%%%%%%%%%%%%%%%%
%%%%%%%%%%%%%%%%%%%%%%%%%%%%%%%%%%%%%%%%%%%%%%%%%%%%%%%%%%%%%%
%%%%%%%%%%%%%%%%%%%%%%%%%%%%%%%%%%%%%%%%%%%%%%%%%%%%%%%%%%%%%%

\section{Empirical Analysis}\label{sec:data}
In this section we present empirical analysis of choice and time data.  We develop an
online experiment to collect choice and time data for pairwise choices.  The first 
goal of this experiment is to establish the existence of a relationship between
choice decisions and response times.  The second goal of this experiment is to show
that the engagement of the user impacts the strength of this relationship.

%%%%%%%%%%%%%%%%%%%%%%%%%%%%%%%%%%%%%%%%%%%%%%
\subsection{Experimental Setup}
We constructed several online polls ranking sets of items or people (we refer to these  simply as items from here on, even if the poll items are people).  The topic of each poll is listed in Table \ref{table:polls} along with the items contained in the poll.   Each poll consisted of five items which were selected to span the range of preferences.  For instance, in the Mathematicians poll, we selected well known and legendary mathematicians such as Carl Gauss and Newton, but we also selected mathematicians of less acclaim such as  Guillaume de l'H\^opital.  For each poll, the subjects are presented pairs of items and for each pair is asked to select the item that they prefer.  Every possible unordered pairing of the items is shown to each subject.  The subjects' choices and response times are recorded for every poll.   
 
\begin{figure}[t]
\centering
\includegraphics[scale=.75]{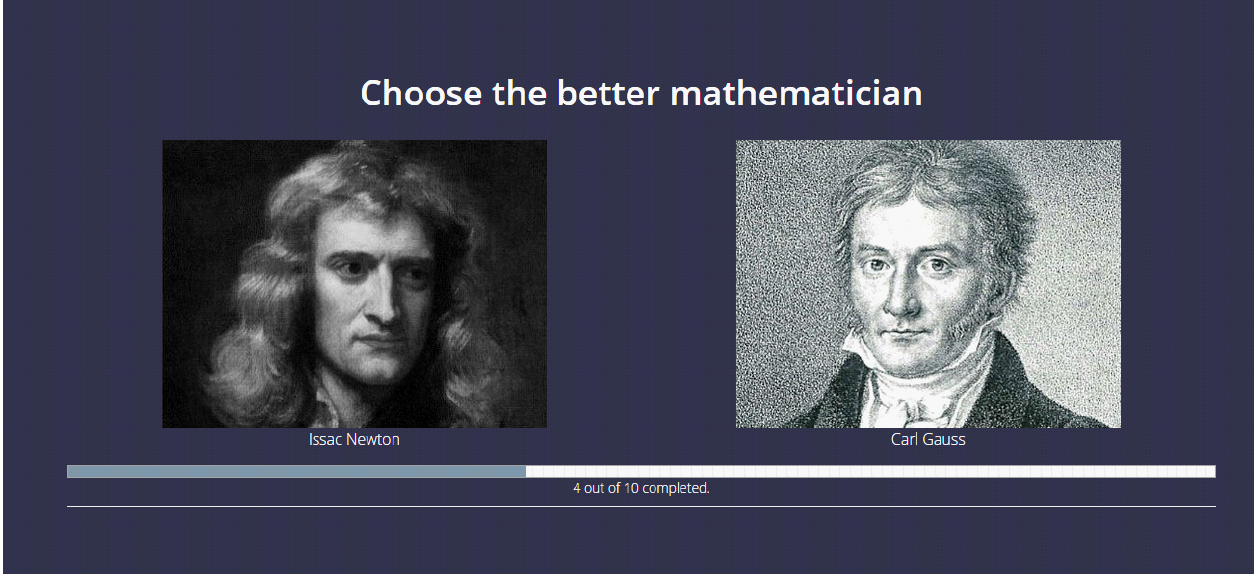}
\caption{Screenshot of the online poll used in the experiment.}
\label{fig:rankbomb}
\end{figure}

 A screenshot of one pair of items from our poll is shown in Figure \ref{fig:rankbomb}.  The ordering of the item pairs is randomized to avoid any spatial biases.  Two checks are done to assess the validity of the data.  Initially, each subject takes a poll where the items are numbers and the subjects are asked to choose the greater number in each pairwise comparison.  We only kept data from users who responded to all of these numerical pairwise comparisons correctly.  Then, before beginning each individual poll,  each subject is shown each item (in random order) and a list of possible names and asked to choose the correct name for the item.  This is done to make sure that the subject has some level of familiarity with the items and will be able to make an informed choice during the poll for all pairs.   For each poll we only use data from subjects that score perfectly on this verification step.

We utilize two sets of subjects for our experiment.  One set consists of students at a university and the other set consists of online Amazon Mechanical Turk (AMT) workers.  The students took the polls while being supervised by a researcher, while the AMT workers took the polls online without any supervision.  The AMT workers took all the polls,
and the students took all the polls except for the Democratic Presidents and Rappers poll.
In total we have data for 42 students and 164 AMT workers.

%\adjustbox{max width=\textwidth}{
\begin{table}
\centering
\begin{tabulary}{1.0\textwidth}{|C|C|C|C|C|C|}
\hline
Poll & Item 1 & Item 2 & Item 3 & Item 4 & Item 5\\
\hline
Mathematicians & Carl Gauss & Isaac Newton & Blaise Pascal & Daniel Bernoulli &  Guillaume de l'H\^opital\\
\hline
Movies & Dances with Wolves & Disaster Movie & The Bank Job & The Godfather & Titanic\\
\hline
Musicians & Bach & Beethoven & Chopin & Mendelssohn & Mozart\\
\hline
Places & Eiffel Tower & Leaning Tower of Pisa & Statue of Liberty & Sydney Opera House & The Pyramids\\
\hline
Democratic Presidents & Barack Obama & Bill Clinton & Jimmy Carter & John F. Kennedy & Lyndon Johnson\\
\hline
Rappers & Eminem & Jay-Z & Notorious B.I.G. & Rakim & Snoop Doggy Dogg\\
\hline 
\end{tabulary} 
\caption{The polls used and corresponding items.}\label{table:polls}
\end{table}

%%%%%%%%
%%

%%%%%%%%%%%%%%%%%%%%%%%%%%%%%%%%%%%%%%%%%%%%%%%%%%%%%%%%%%%%%%%%%%%%%%%%%%%%%%%%%%%%%%%%%%%%%%%%%%%%%%%%%%%%%%%%
%%%%%%%%%%%%%%%%%%%%%%%%%%%%%%%%%%%%%%%%%%%%%%%%%%%%%%%%%%%%%%%%%%%%%%%%%%%%%%%%%%%%%%%%%%%%%%%%%%%%%%%%%%%%%%%%
\subsection{Mean and Correlation of Choice and Time Data}
\begin{figure}[t]
\centering
\includegraphics[scale=1]{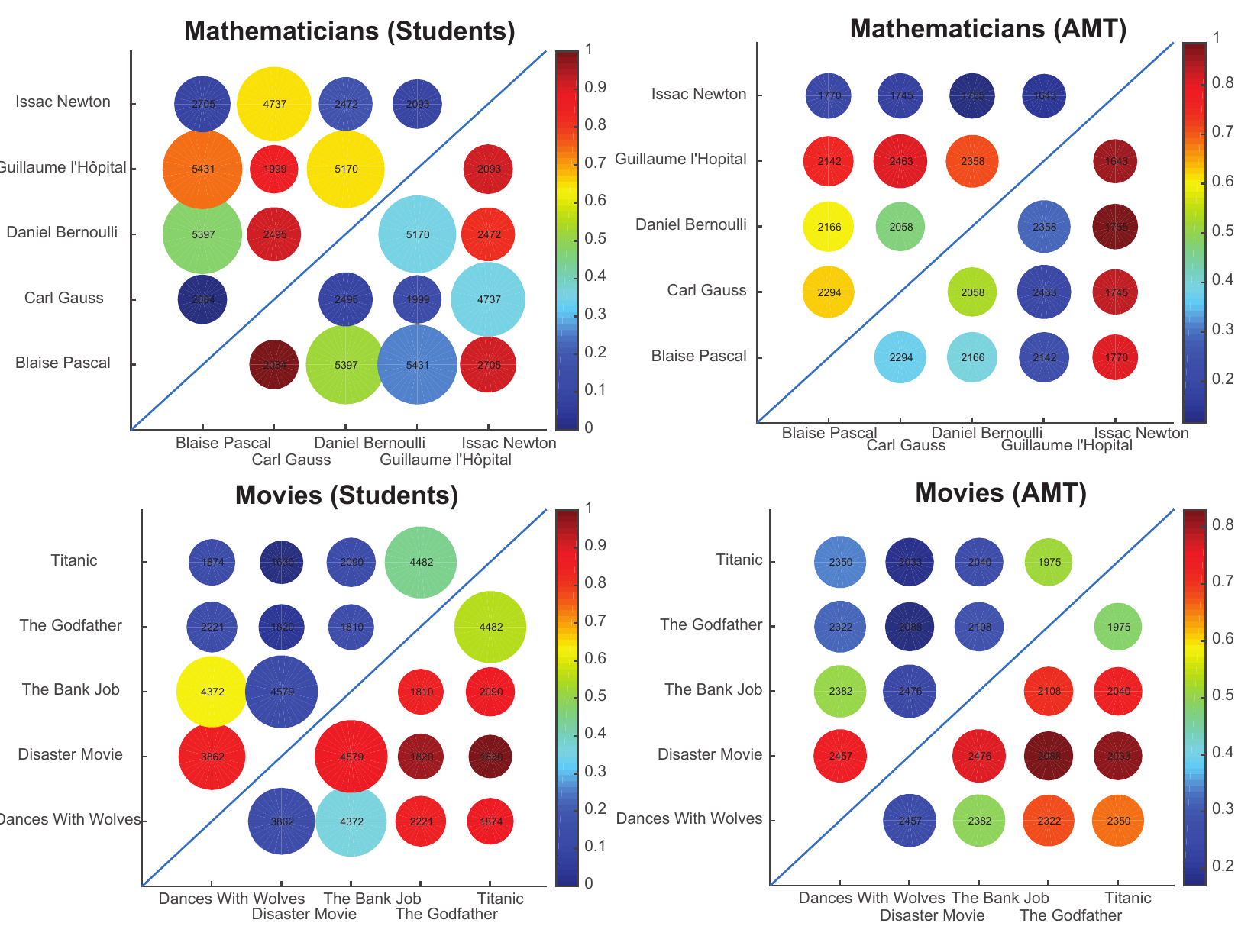}
\caption{Bubble plots for (top left) Mathematicians (students), (top right) Mathematicians (AMT), 
(bottom left) Movies  (students), and (bottom right) Movies  (AMT).
The bubble size is proportional to the average response time and each bubble is labeled with the average response time (in milliseconds).  The bubble color indicates the fraction of time the item on the horizontal axis is chosen when compared to the item on the vertical axis.  The corresponding color map is shown on the right of each bubble plot.}
\label{fig:bubble_plots}
\end{figure}

We begin our exploratory analysis by looking at some simple statistics of the data to understand the connection between choice and time.  To visualize the data for each poll we use bubble plots which encode the timing and choice data for each pair of items shown to the subjects.  We show the bubble plots for the student and AMT subjects for the Mathematicians and Movies polls in Figure \ref{fig:bubble_plots}.   The bubble color indicates the fraction of time the item on the horizontal axis is chosen when compared to the item on the vertical axis.  The size of each bubble is proportional to the average response time for each item, which is also displayed in each bubble.   The bubble plots allow us to see the choice and time connection.  We observe that when two items are of different qualities, i.e. one is strongly preferred to the other, the total amount of time taken to decide between them is very short. For instance, in the Mathematicians poll, for the students, it takes roughly two seconds on average to decide between Gauss and L'H\^opital, and in this pairing Gauss is chosen approximately 89\% of the time.  When two items are of similar qualities, the total amount of time taken to decide between them takes longer. In the Movies poll, for the students it takes 4.5 seconds on average to decide between Titanic and The Godfather, and in this pairing the Godfather is chosen approximately 41\% of the time.

\begin{figure}[t]
\centering
\includegraphics[scale=1]{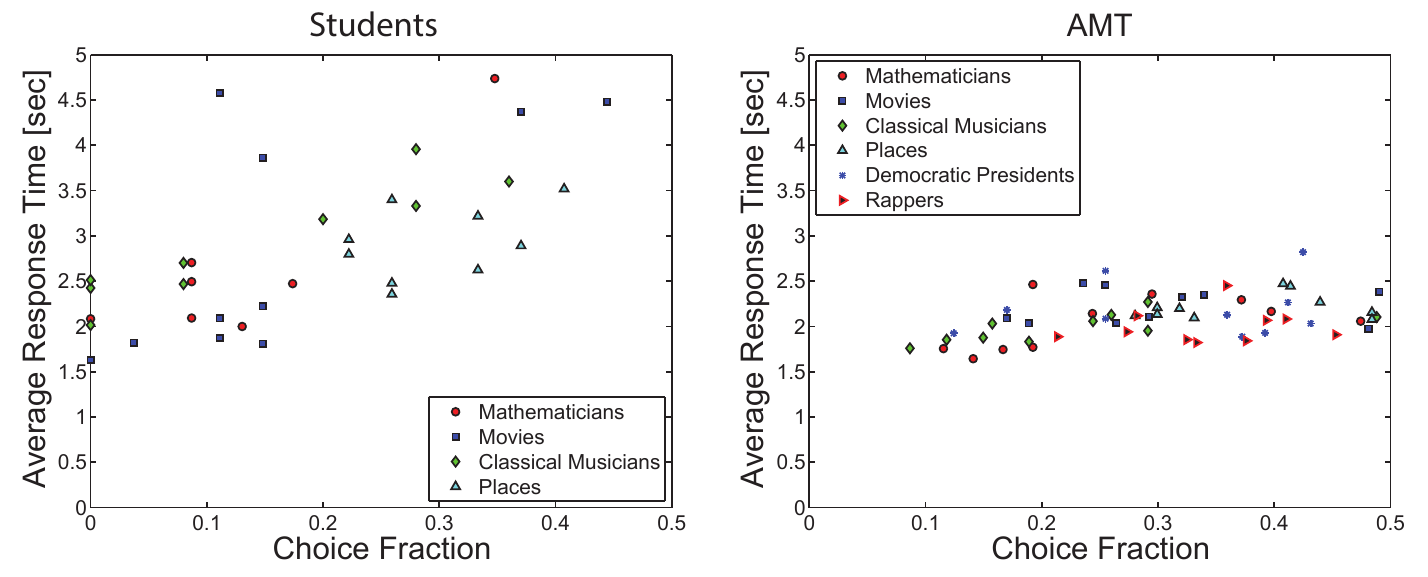}
\caption{Plot of choice fraction versus mean response time for student and AMT subjects.}
\label{fig:choice_vs_time}
\end{figure}

To make this connection between choice and time even more evident, we plot in Figure \ref{fig:choice_vs_time} the average response time versus the choice fraction (the fraction of times the less popular item is chosen) for each item pair for  all polls and subjects.  This restricts the choice fractions to the interval $[0,1/2]$.  The data for students and AMT subjects are plotted separately.   As can be seen there is a clear correlation between the choice and time data for both populations.  The correlation coefficient between the choice fractions and average response times is 0.84 (p-value $<10^{-6}$) for the students and 0.55 (p-value  $<10^{-6}$) for the AMT subjects.  This indicates that there is a clear correlation between choice and time in this data, but the strength of the correlation is different for the two populations.  
The students have a higher correlation between their choice and time data than the AMT subjects. 

\begin{figure}[t]
\centering
\includegraphics[scale=1]{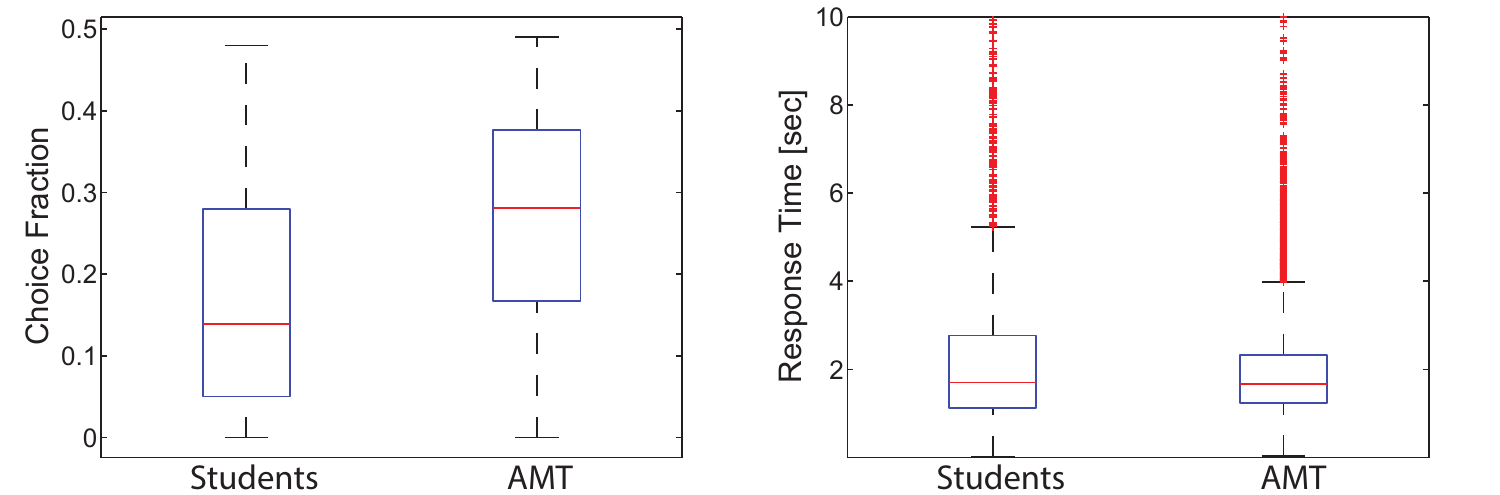}
\caption{Boxplots of the choice fractions (left) and response times (right) for the student and AMT subjects.}
\label{fig:boxplots}
\end{figure}

To further analyze the difference between the choice and time data for these two populations, we look at some aggregate statistics of the data.  In Table \ref{table:ug_vs_mt} we list the mean choice fraction and response time (averaged across all polls and users) for the two populations.  We see that on average, the response times are shorter for the AMT subjects (2.11 seconds) than for the students (3.18 seconds).  For the choice fractions, a value of 0.5 would indicate random choice decisions.
 The mean choice fractions are closer to random for the AMT subjects (0.21) than the students (0.30).  We see here that the AMT subjects have  lower engagement for the poll task.  They are most likely thinking less about their choices, and are therefore able to react more quickly.  The students seem to be more careful in their choices, resulting in slower response times, which may be due to the fact that they were supervised by the researcher.  This more careful choice decision also would explain the difference in mean choice fraction for the two populations.  Recall that the choice fractions are upper bounded by 0.5, which corresponds to completely random choices.   The lower mean in the choice fraction for the students comes from the fact that they give some items a lower choice fraction because they are more careful about their choices
and choose certain items much less often than the AMT subjects who are choosing more randomly.  This effect can also be seen in Figure \ref{fig:boxplots} where we show boxplots of the choice fraction and response time for the two populations.  The median choice fraction is lower and the variance is slightly higher for the students, which  agrees with the engagement hypothesis.  For the response times, we see that the students have a higher median
and variance.  This also agrees with our engagement hypothesis.

Our exploratory analysis suggests that time is not only connected to choice, but also to the level of engagement
with the choice task.  By not accounting for  user engagement in a choice model, one could end up
biasing estimates of item preferences.  A choice model which captures these features could allow one to
discern the interest of users in the choice decisions they make, allowing for more accurate
measurement of true item preferences.  This motivates our construction of a choice engagement time 
model in Section \ref{sec:cet_model}.

%data for table
%Undergraduates:\n\tE[P] = 0.17777:\n\t E[T] = 2.5935
%Mechanicial Turk:\n\tE[P] = 0.26293:\n\t E[T] = 2.0023
\begin{table}
\centering
\begin{tabulary}{0.70\textwidth}{|C|C|C|C|}
\hline
Subjects & Mean choice fraction & Mean response time [seconds]& Correlation coefficient of choice fraction and mean response time \\
\hline
UG         &0.21 & 3.18  & 0.74 (p-value $<10^{-5}$) \\
\hline
AMT &0.30 & 2.11  & 0.35 (p-value $0.006$) \\
\hline 
\end{tabulary} 
\caption{Choice and time statistics for student and AMT subjects.}\label{table:ug_vs_mt}
\end{table}

%%%%%%%%%%%%%%%%%%%%%%%%%%%%%%%%%%%%%%%%%%%%%%%%%%%%%%%%%%%%%%%%%%%%%%%%%%%%%%%%%%%%%%%%%%%%%%%%%%%%%%%%%
%%%%%%%%%%%%%%%%%%%%%%%%%%%%%%%%%%%%%%%%%%%%%%%%%%%%%%%%%%%%%%%%%%%%%%%%%%%%%%%%%%%%%%%%%%%%%%%%%%%%%%%%%
\subsection{Comparison of Conditional Response Time Distributions}
We next look at the symmetry of the timing data.  In particular,
we would like to know if the choice made affects the response time
for a given pair of items.  For instance, if comparing
 items $a$ and $b$, is the  response time distribution conditional on $a$ being chosen different from
 the reaction time distribution conditional  on $b$ being chosen? 
For each pair of items in a poll, our null hypothesis is that these conditional distributions
are equivalent and our alternative hypothesis is that they come from distributions with different medians.
We test each subject population separately, as there may be different effects due to the circumstances in which
the polls were taken.  

For each pair of items in each poll we conducted a two-tailed Wilcox Mann-Whitney non-parametric rank sum test.
This tests whether or not the two groups (choosing item $a$ or choosing item $b$) came from distributions with different
medians.  Because for each subject population we are testing multiple hypotheses (one for each item pair),
to assess significance we must employ the Bonferroni correction \citep{dunn1961multiple}.  This corrections states that
when testing $m$ hypotheses, to keep the family wide error rate (probability of making at least one
type I error) at a level $\alpha$, one must test each individual
hypothesis at a level $\alpha/m$.  If any individual null hypothesis has p-value less than $\alpha/m$,
then one can reject it.  
This corrects for the fact that with multiple hypotheses, the chance of a rare
event increases and the probability of incorrectly rejecting a null hypothesis
increases.

Each item pair considered  corresponds to a hypothesis being tested.
We can only consider item pairs where both items have at least one vote.  This restriction gives us 33 item pairs
for the student population and 60 item pairs for the  and AMT population.  
We test the null hypothesis that the medians are equal at a 5\% level 
 for each subject population separately.  After applying
 the Bonferroni correction (with $m=33$ for the students and $m=60$ for the
AMT subjects) we find that
we cannot reject the null hypothesis for both populations.  All item
pairs have p-values below the corresponding Bonferroni corrected threshold. This suggests that there
may not be any difference in the distribution of the response times conditioned on the item chosen.
This finding will impact the construction of our choice engagement time model in Section \ref{sec:cet_model}.

%%%%%%%%%%%%%%%%%%%%%%%%%%%%%%%%%%%%%%%%%%%%%%%%%%%%%%%%%%%%%%%%%%
\section{Previous Choice Time Models}\label{sec:previous_models}
Our empirical analysis has shown a relationship between choice decisions and response times.
There have been different models developed to explain this phenomenon
by directly modeling the mental decision process.  These models were initially
developed to explain the connection not between choice and time, but between accuracy 
and time in certain cognitive tasks.  However, 
choice and accuracy are naturally related if one views choosing the preferred item  
as trying to accurately select the better item.

In this section we review the main choice time models
from the psychology literature: the Poisson counter model \citep{pike1973response,townsend1983stochastic} and the
drift diffusion model \citep{ratcliff1978theory}.  For each model we review the psychological assumptions underlying the model and also analyze the various predictions they make.  
We show that these models each possess different shortcomings.  They either do not align with our empirical data or they provide computational challenges for model estimation.  Our model builds upon the main
properties of these models but is also relatively simple both for interpretation
and estimation.

Throughout this section we assume the choice decision is made between two items $a$ and $b$.
For a given model, we let $p$ be the probability that item $a$ is chosen (and therefore $1-p$ is the probability that $b$ is chosen),
$\mu$ be the average response time, and $\mu_a$ and $\mu_b$ be the average response times
conditioned on $a$ or $b$ being chosen.  We will investigate the predictions
made for these parameters for the Poisson counter and drift diffusion models.

%%%%%%%%%%%%%%%%%%%%%%%%%%%%%%%%%%
\subsection{Poisson Counter Model}\label{pc_model}
The Poisson counter model is an accumulator based choice time model where the information accumulation process
is modeled as a Poisson process \citep{pike1973response, townsend1983stochastic}.  In the model, there are two information accumulation processes $N_a(t)$ and $N_b(t)$ for the two possible responses $a$ and $b$ with rates 
$\alpha$ and $\beta$, respectively.  These  processes are modeled as Poisson processes.    They accumulate information independently and in parallel.  Each process has its own threshold $K_a$ or $K_b$, respectively.  The resulting choice made corresponds to the  process that reaches its threshold first.  The more preferred an item is, the larger the
rate of its corresponding information accumulation process.   

\begin{figure}
\centering
\includegraphics[scale=0.4]{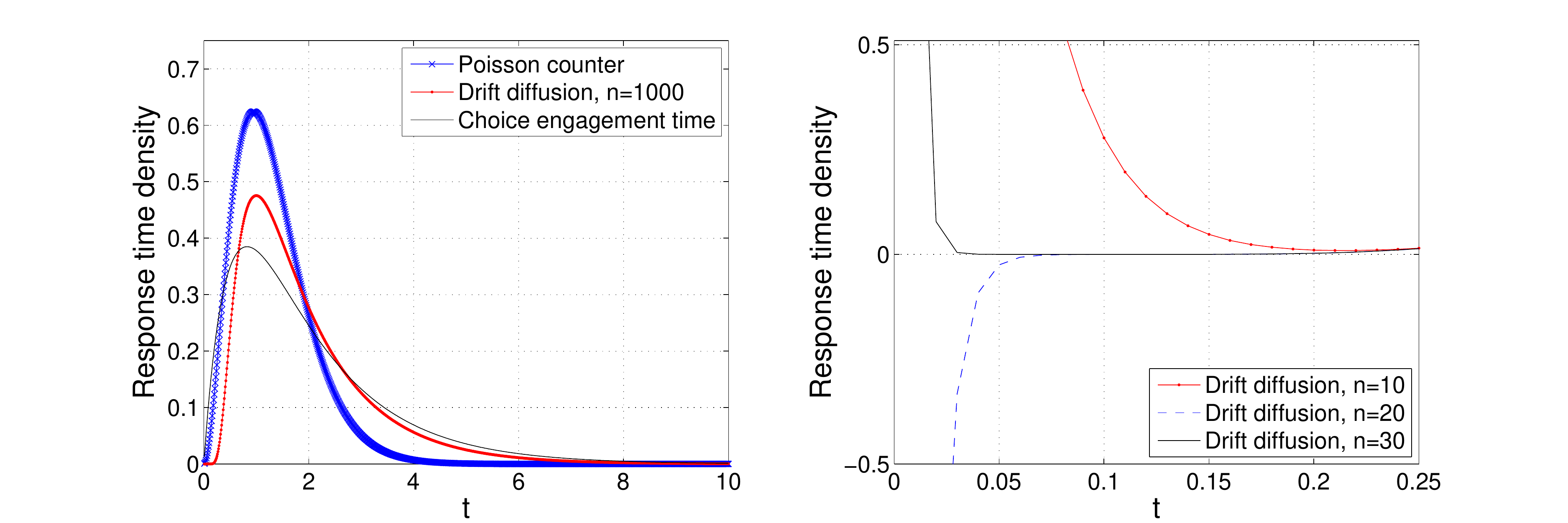}
\caption{(left) Plot of the response time density for the Poisson counter, drift diffusion, and choice engagement time models.
The Poisson counter model has $K_a=K_b = 3$, $\alpha = 3$, and $\beta=1$.  The drift diffusion model
has $z=2$, $K = 4$, $d = 1$, $\sigma^2=1$, and $n=1000$ terms in the summation..  For the choice engagement time model, the hypoexponential density parameters are $a=1.5$ and $b=0.5$. (right) Plot of the response time density of the drift diffusion model 
 with $z=2$, $K = 4$, $d = 1$,  $\sigma^2=1$, and $n=10, 20, 30$ terms in the summation.  }
\label{fig:hitting_time_density}
\end{figure}

\begin{figure}
\centering
\includegraphics[scale=.5]{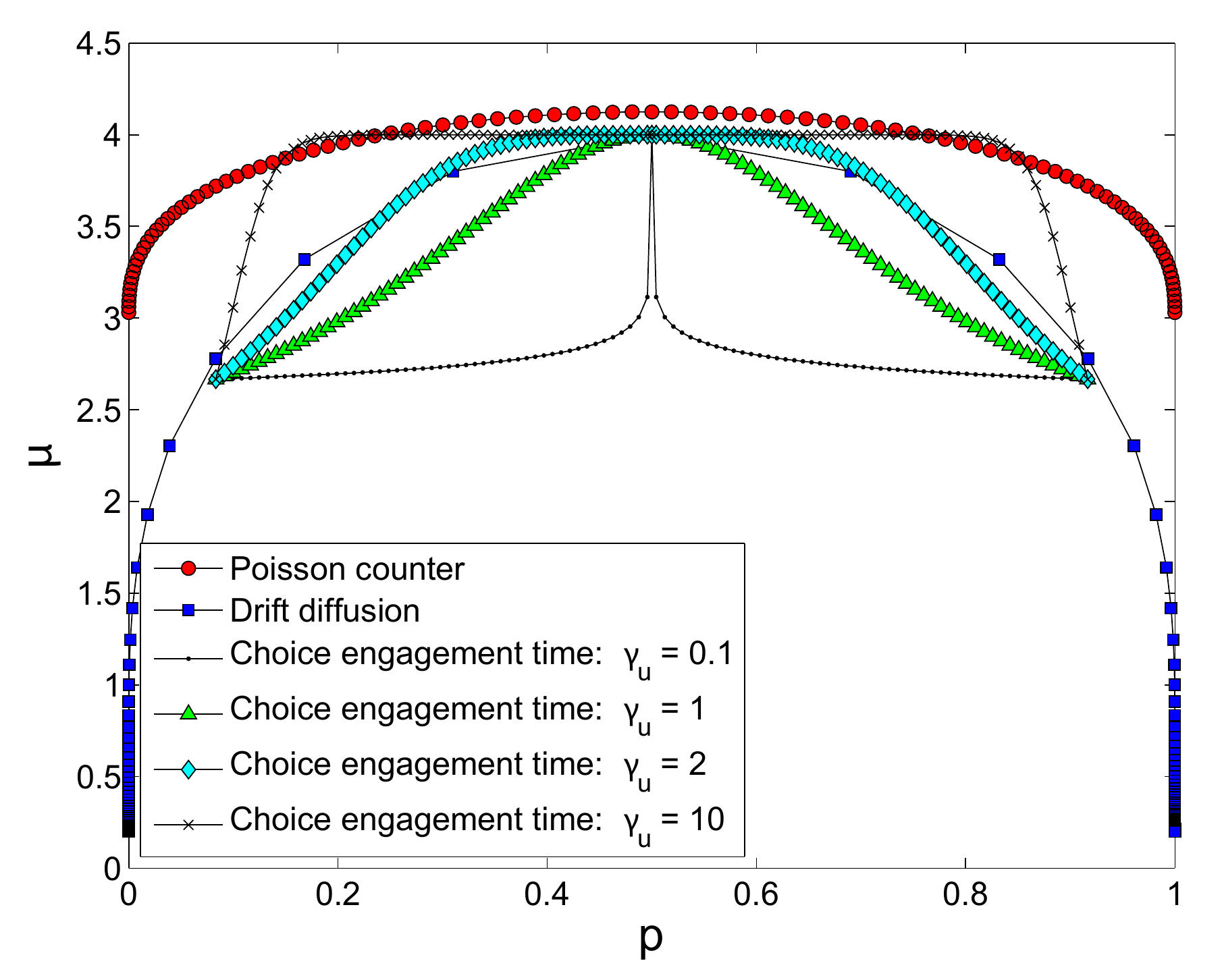}
\caption{Plot of the mean response time $\mu$ versus choice probability $p$ for the Poisson counter, drift diffusion, and choice engagement time models.  For the Poisson counter model $K_a=K_b=3$, $\beta=1-\alpha$, and $\alpha$ is swept from zero to one.  For the drift diffusion model $K=4$, $z=2$, $\sigma^2=1$, and $d$ is swept from zero to -10 to 10.  For the choice engagement time model $\tau_u =2$, $A_u = 1$, $\epsilon_u = 0.1$, $\rho_u = 0.4$, $\gamma_u=0.1,0.5,2,10$, and the normalized item utility is swept from zero to one.}
\label{fig:mu_vs_p}
\end{figure}

 For a Poisson process with rate $\lambda$,
the time to have $K$ arrivals has an Erlang distribution with parameters  $K$ and $\lambda$  and mean $K/\lambda$.  We denote the density of an Erlang random variable as 
\begin{align}
	f_E(t;K,\lambda)& = \frac{\lambda^{K} t^{K-1} e^{-\lambda t}}{(K - 1)!}.\label{eq:erlang_pdf}
\end{align}
and its corresponding cumulative distribution function (CDF) as
\begin{align}
	F_E(t;K,\lambda)& = 1-\sum_{n=0}^{K-1}\frac{\lambda^{K} t^{K} e^{-\lambda t}}{n!}.\label{eq:erlang_cdf}
\end{align}
A choice of $a$ will be made with response time $t$ if at time $t$ $N_a(t)$ has arrival $K_a$
and $N_b(t)$ has not had arrival $K_b$ yet.
With our notation, we can then write the joint likelihood of observing a choice $c\in\curly{a,b}$ and corresponding response time $t$ as
\begin{align}
	f(c,t) & = \begin{cases}
	f_E(t;K_a,\alpha)\paranth{1-F_E(t;K_b,\beta)},~~~ c=a\\
	f_E(t;K_b,\beta)\paranth{1-F_E(t;K_a,\alpha)},~~~ c=b.
	\end{cases}\label{eq:likelihood_poisson}
\end{align}
Using this likelihood function we can obtain useful properties about the Poisson
counter model.  For instance, the probability density function (PDF) of a response time $t$ marginalized over the
choice made is simply
\begin{align}
	f(t) & = f_E(t;K_a,\alpha)\paranth{1-F_E(t;K_b,\beta)}+f_E(t;K_b,\beta)\paranth{1-F_E(t;K_a,\alpha)}.
\end{align}
We plot an example of this density in Figure \ref{fig:hitting_time_density}.  We see two features from this plot.  First, the density
has a mode.  Second, the tail of the density decays exponentially fast.  We will come back to these properties later when
we propose our choice engagement time model.  In addition to the response time density, we have the following results for the Poisson counter model, which are proved in the appendix:
%%%%%%%%%%%%%%%
\begin{thm}\label{thm:poisson}
For the Poisson counter model with information accumulation processes with rates $\alpha$ and $\beta$ and thresholds $K_a$ and $K_b$ and for items $a$ and $b$, respectively, we have that
\begin{align}
		p & = I_{\alpha/(\alpha+\beta)}(K_a,K_b)\\
		\mu_a & =  \frac{K_a}{p\alpha}I_{\alpha/(\alpha+\beta)}(K_a+1,K_b)\\
		\mu_b & =  \frac{K_b}{(1-p)\beta}I_{\beta/(\alpha+\beta)}(K_b+1,K_a)\\
    \mu & =  \frac{K_a}{\alpha}I_{\alpha/(\alpha+\beta)}(K_a+1,K_b)+
	          \frac{K_b}{\beta}I_{\beta/(\alpha+\beta)}(K_b+1,K_a).
\end{align}

where we have defined the regularized incomplete beta function as
\begin{align}
	I_q(a,b) & = \paranth{\int_0^1t^{a-1}(1-t)^{b-1}dt}^{-1}\paranth{\int_0^qt^{a-1}(1-t)^{b-1}dt}\label{eq:beta_inc}.
\end{align}
\end{thm}
A natural assumption to make is that the information thresholds are equal for each item.  
  This would correspond to a cognitive process where an item is selected when the accumulated evidence reaches a single fixed threshold.  We set $\beta=1-\alpha$ and $K_a=K_b$ and 
 plot the resulting relationship between $p$ and $\mu$ as a function of $\alpha$  in Figure \ref{fig:mu_vs_p}.  
We see that $\mu$ is symmetric about $p=1/2$ and reaches its peak here.  When the rates are unequal and one
item is more preferred than the other,  the
mean response time decreases, similar to what we
observed in our empirical data.  

There is one property of the Poisson counter model that does not align with our empirical observations.
It can easily be checked that even with symmetric thresholds,
the conditional means $\mu_a$ and $\mu_b$ will be different.  This lack of symmetry goes against our empirical
observations.  Therefore,  the Poisson counter model is not the ideal model for two item 
choice decisions and response times.

%%%%%%%%%%%%%%%%%%%%%%%%%%%%%%%%%%%%%%%%%%%%%%%
\subsection{Drift Diffusion Model}\label{sec:ddm_model}
Another popular model for choice and response times is the
 drift diffusion model \citep{ratcliff1978theory}.  Similar to the Poisson counter model, in this model when a person has to choose between items $a$ and $b$, an information accumulation process begins.  However, in this case the  process is a one dimensional Brownian motion with variance $\sigma^2$, drift $d$, and initial value $z>0$.  The process evolves until it hits
one of two thresholds located at $0$ and $K>z$.  If $K$ is hit first, item $a$ is chosen, otherwise if 0 is hit first item $b$ is chosen.  If  $a$ is more preferred than $b$, then $d$ will be positive with large magnitude, whereas if  $b$ is more preferred than $a$, then $d$ will be negative with large magnitude.  If $a$ and $b$ are similar, then $d$ will be close to zero.

We draw upon results from the theory of Brownian motion to obtain the likelihood of the choice and response time data.
We make use of the following result for hitting times of a Brownian motion.
\begin{lem}[\citep{feller1959introduction} pp. 327]
Let $X(t)$ be a Brownian motion with drift $d\neq 0$, variance $\sigma^2$, and initial value $X(0)= z>0$.  Let there be
boundaries at $K>z$ and 0. Let $u(t,z,K,d,\sigma^2)$ be the density of the first time to hit 0 before hitting $K$.  Then
\begin{align}
	u(t,z,K,d,\sigma^2) & = \frac{2\pi}{K^2}\exp\paranth{-\frac{dz}{\sigma^2}}\sum_{n=1}^{\infty}n\exp\paranth{-\paranth{\frac{\sigma^2\pi^2n^2}{K^2}+\frac{d^2}{\sigma^2}}t}\sin\paranth{\frac{\pi zn}{K}}
\end{align}
\end{lem}
We note that to obtain the density of the hitting time of $K$ before 0 is hit, we simply replace $d$ by $-d$ and $z$ by $K-z$.
With this we can easily write down the likelihood of choice and time data $c$ and $t$ as
\begin{align}
	f(c,t) = \begin{cases}
	        u(t,z,K,d,\sigma^2), ~~~ c=a\\
					u(t,K-z,a,-d,\sigma^2), ~~~ c=b.
					\end{cases}
\end{align}\label{eq:likelihood_ddm}
and the density of the response time  marginalized over choice is
\begin{align}
	f(t) & = u(t,z,K,d,\sigma^2)+u(t,K-z,K,-d,\sigma^2).
\end{align}
We plot an example of this density in Figure \ref{fig:hitting_time_density}.  The response time density for the drift diffusion model is qualitatively  similar to the Poisson counter model.  For instance, it
has the same uni-modal and exponential tail properties of the Poisson counter model.  However, the drift diffusion model density involves
an infinite sum and there can be convergence issues for small values of $t$ if not enough terms are kept in the sum, as can be seen in the figure.  This can create challenges for estimating this model.  Our proposed choice engagement time model
will retain the qualitative properties of the response time densities, but be simple enough to enable straightforward
estimation.

We can utilize standard martingale theory to obtain further properties of the drift diffusion model.  We have the following results, with proof in the appendix.
%%%%%%%%%%%%%%%
\begin{thm}\label{thm:ddm}
For the drift diffusion model with inital value $z>0$, thresholds $K>z$ and $0$ for items $a$ and $b$, drift $d\neq 0$, and variance $\sigma^2$, we have that
\begin{align}
		p & =  \frac{1-e^{-2dz/\sigma^2}}{1-e^{-2dK/\sigma^2}}\\
			  \mu_a & = \frac{2K^2e^{\frac{d(K-z)}{\sigma^2}}}{p\pi^3\sigma^2}\sum_{n=1}^{\infty} \frac{n\sin\paranth{\frac{\pi (K-z) n}{K}}}{\paranth{n^2+\paranth{\frac{dK}{\pi\sigma^2}}^2}^2} \\
		\mu_b & =\frac{2K^2e^{-\frac{dz}{\sigma^2}}}{(1-p)\pi^3\sigma^2}\sum_{n=1}^{\infty} \frac{n\sin\paranth{\frac{\pi z n}{K}}}{\paranth{n^2+\paranth{\frac{dK}{\pi\sigma^2}}^2}^2}   \\ 
    \mu & =  \frac{pK-z}{d}.
\end{align}
\end{thm}
%%%%%%%%%%%%%%%%%%%%%%%%%%%%%%%%
 As with the Poisson counter model, to gain insight into the model we focus on the symmetric case where $z=K/2$.
In Figure \ref{fig:mu_vs_p} we show the resulting relationship between $\mu$ and $p$. 
The drift diffusion model has a natural symmetry in $\mu$ with respect to $p$. 
Response times are short for large $|d|$, which corresponds to a strong preference for one of the items.
This is similar to what is seen for the Poisson counter model.   However, there is a fundamental difference
between the two models concerning the conditional mean response times.  We have the following result, which is proved
in the appendix:
%%%
 \begin{lem}\label{lem:ddm_equal}
For the drift diffusion model with inital value $z>0$, thresholds $2z$ and $0$ for items $a$ and $b$, drift $d\neq 0$, and variance $\sigma^2$, we have that $\mu_a=\mu_b$.
\end{lem} 
As can be seen, unlike in the Poisson counter model, the conditional mean response times
are symmetric under symmetric thresholds.  This is in alignment with our empirical
observations.  The drift diffusion model is therefore a better model for our choice and
response time data.  However, it still suffers from estimation challenges due to the complexity
of its likelihood function.  We overcome this challenge with a simpler choice engagement
time model which we propose next.

%%%%%%%%%%%%%%%%%%%%%%%%%%%%%%%%%%%%%%%%%%%%%%%
\section{Choice Engagement Time Model}\label{sec:cet_model}
Our empirical analysis revealed a correlation between response time and choice  and also showed the impact
of user engagement on choice decisions. In this section we propose a choice engagement time model which incorporates the choice and time relationship, but also captures the user engagement phenomenon.
Our model will keep the uni-modality and exponential tail of the response time density
we observed for the Poisson counter and drift diffusion models.  It will also possess symmetric conditional mean response times as we saw in our empirical data and in the drift diffusion model.   In addition, we also want the likelihood function of the model to have a simple functional form to allow for tractable estimation.  We will focus on pair-wise comparisons.  That is, each choice decision will involve the user selecting the preferred item from an offer set of two items.

\subsection{Basic Model}
We begin by assuming that each user $u$ has a set of user specific parameters 
$\Theta_u = \curly{\epsilon_u,A_u,\tau_u,\gamma_u,\rho_u}$, all of which are non-negative.  These parameters will govern the timing and choice behavior of the user.  We assume the choice and time data to be independent
 conditioned on the user model parameters.  For the choice data, we use a simple modification of 
the common Bradley-Terry  choice model \citep{bradley1952rank}.  We assume there is a set $\mathcal M$ of $M$ items with
 utilities $\curly{w_1,w_2,...,w_M}$ such that $w_i\geq 0 $ for $1\leq i \leq M$.  To avoid
identifiably issues, we assume the normalization  $\sum_{i=1}^M w_i = 1$. 
Let $C_{uij}^k$ be the choice made in the $k$th comparison of items $i$ and $j$ by user $u$.  $C_{uij}^k$ is one if item $i$ chosen, and zero if $j$ is chosen.  We model $C_{uij}^k$ as a Bernoulli random variable which is one with probability $p_{uij}$.
Our model for $p_{uij}$ incorporates the item utilities as in the traditional Bradley-Terry model, but also adds in
user engagement through the parameter $\epsilon_u$  as follows.  If items $i$ and $j$ are offered to user $u$, then let the normalized utility of items $i$ and $j$ be $w'_i=w_i/(w_i+w_j)$ and  $w'_j=1-w'_i$, respectively. We define $w_{ui} = w'_i+\epsilon_u$ and $w_{uj} = w'_j+\epsilon_u$ as user specific utilities for an offer set $\curly{i,j}$.  For very large $\epsilon_u$, the user specific utilities will be approximately equal to $\epsilon_u$ for all items.   The effect of this is to reduce the difference in the items' utility for the the user.    This corresponds to a user with low engagement  who sees all items as equal and chooses between them in a  random manner because it does not make any difference to him which one is chosen.  We let the choice probabilities be the same as in the Bradley-Terry model, but with the user specific utilities used instead of the underlying utilities, giving
\begin{align}
p_{uij} &= \frac{w_{ui}}{w_{ui}+w_{uj}}\nonumber\\
&=\frac{\frac{w_{i}}{w_{i}+w_{j}}+\epsilon_u}{1+2\epsilon_u}\label{eq:p}.
\end{align}
%%%
The choice probability is bounded by $\epsilon_u\paranth{1+2\epsilon_u}^{-1}$ and              
$\paranth{1+\epsilon_u}\paranth{1+2\epsilon_u}^{-1}$.
For $\epsilon_u$ near zero the choice probability reduces to the Bradley-Terry model, 
but in the limit of large $\epsilon_u$, the user will choose either item with probability $1/2$,  
thereby exhibiting low engagement.  Larger $\epsilon_u$  also results in less variation in the choice probability across item pairs.

  For each choice decision $C_{uij}^k$ we also
have the corresponding response time $T_{uij}^k$.  We model the response time as the sum of a latency time $L_{uij}^k$
and a decision time $D_{uij}^k$.  The latency time models how long it takes the user to view the items
and recognize what they are.  We model this as an exponential random variable with mean value $\tau_{u}$.
The decision time models how long it takes the user to decide which item to choose once he recognizes
the items.  Similar to the latency time, we also model this as an exponential random variable.  However,
for item utilities $w_i$ and $w_j$ the mean decision time $\delta_{uij}$ is given by
\begin{align}
	\delta_{uij} & = \frac{A_u}{\paranth{\frac{w_i-w_j}{w_i+w_j}}^{2\gamma_u}+\epsilon_u+\rho_u},
\end{align}
 and the resulting mean response time is given by
\begin{align}
	\mu_{uij}  & = \tau_u +\delta_{uij}\nonumber\\
	& =  \tau_u+\frac{A_u}{\paranth{\frac{w_i-w_j}{w_i+w_j}}^{2\gamma_u}+\epsilon_u+\rho_u}.\label{eq:muT}
\end{align}
The absolute scale of the time is captured by $\tau_u$ and $A_u$, both of which have units of time.  The parameters $\epsilon_u$ and $\gamma_u$ capture the user engagement.  While $\epsilon_u$ relates uniformity of choice probability to fast response times,
$\gamma_u$ measures the sensitivity of the response times to the item utilities.  The term $\rho_u$ is simply a regularizer that controls the mean decision time for users with high engagement.  The mean response time is bounded by
	$\tau_u+A_u\paranth{1+\epsilon_u+\rho_u}^{-1}$ and 
	$\tau_u+A_u\paranth{\epsilon_u+\rho_u}^{-1}$.
The minimum value is achieved for $w_i$ or $w_j$ equal to zero, and the maximum value is achieved for $w_i=w_j$.  For a user with maximum engagement ( $\epsilon_u=0$), the mean response time is upper bounded by $A_u/\rho_u$.  This is how the term $\rho_u$ acts to regularize the mean response time.  For large $\epsilon_u$, the ratio of the maximum to minimum mean response time approaches one.  Therefore, for very low engagement users, there is reduced variation in the mean response time across items.  This models the low engagement of the user because his response time does not depend upon the items.  

The upper and lower bounds of the mean response time are set by $\tau_u,A_u,\epsilon_u$ and $\rho_u$.  However, the shape of the mean  response time as a function of the item utilities is governed by $\gamma_u$.  For large $\gamma_u$,  $\mu_{uij}$ is very sensitive to item utility for choice probabilities near zero or one.  For small $\gamma_u$, $\mu_{uij}$ is very sensitive to item utility for choice probabilities near one half.
We show the relationship between the choice probability and mean  response time in Figure \ref{fig:mu_vs_p}.
 As can be seen, by varying $\gamma_u$ we can adjust the shape of the curve. We have more flexibility with the choice engagement time model than with the drift diffusion model, but still maintain the basic symmetry of mean  response time with respect to choice probability.

The response time is the sum of two exponential random variables and therefore  it has a hypoexponential distribution.  Formally, let $T$ be a random variable that is the sum of two exponential random variables with means $a$ and $b$.  Then $T$ is a hypoexponential random variable and we will use the convention that it is characterized by parameters $a$ and $b$.  It has mean $a+b$ and density 
\begin{align}\label{eq:hypoexponential}
	g(t;a,b) & =\begin{cases}
	         (b-a)^{-1}\paranth{e^{-t/b}-e^{-t/a}},~~~ a\neq b\\
					 a^{-2}te^{-t/a},~~~ a=b.
					\end{cases}
\end{align}
We show an example of  the density of a hypoexponential response time in Figure \ref{fig:hitting_time_density}.  
  As can be seen, the shape of this density mimics that of the Poisson counter and drift diffusion models (exponential tail and  
	a single mode), but with a much simpler functional form, allowing for more tractable model estimation.

%%%%%%%%%%%%%%%%%%%%%%%%%%%%%%%%%%%%%%%%%%%%%%%%%%%%%%%%%%%%%%%%%%%%%%
\subsection{Engagement with a No-Purchase Option}\label{sec:swipe_right}
We have modeled low engagement as an increase in the randomness of user choice.  However, this is not the only way engagement levels are manifested in choice and time data.  For instance, we saw earlier in Tinder that
low engagement users select ``yes'' for every match shown.
This type of low engagement behavior where choices are biased against the no-purchase option
can be modeled in our framework by giving every item a higher utility relative to the no-purchase option.
We set the  utility of the no-purchase option to $w_0$. We use our existing model with every offer set containing an item $i$ and the no-purchase option.  As before, we define the normalized item utilities $w'_i = w_i/(w_i+w_0)$ and $w'_0 = 1-w'_i$.   The user specific utility for item $i$ is $w_{ui} = w'_i+\epsilon_u$, as in the original choice engagement time model.  However, for the no-purchase option we set $w_{u0} = w'_0$.  The no-purchase option does not gain any utility from the user's engagement.      This results in a  choice probability given by
\begin{align}
p_{ui0} & = \frac{w_{ui}}{w_{ui}+w_{u0}}\nonumber\\
&= \frac{\frac{w_{i}}{w_{i}+w_{0}}+\epsilon_u}{1+\epsilon_u}\label{eq:ptinder}.
\end{align}
This choice probability will approach one for large $\epsilon_u$, in contrast to the original choice engagement time model where the choice probability approached 1/2 for large $\epsilon_u$.  Therefore, for very low engagement users, 
each items is preferred to the no-purchase option.  

In addition to preferring each item, low engagement could also result in users preferring the no-purchase option.  This would occur if for some reason the users found it beneficial to reject all items.  This behavior could also easily be modeled  by adding $\epsilon_u$ to the normalized utility of the no-purchase option and leaving all other item utilities unchanged.  Therefore, the choice  engagement time model can capture different user engagement behaviors that may occur when the offer set is a single item and a no-purchase option.

%

%%%%%%%%%%%%%%%%%%%%%%%%%%%%%%%%%%%%%%%%%%%%%%%%%%%%%%%%%%%%%%%%%%%%%%%%%%%%%%%%%%%%%%%%%%%%%%%%%%%%%%%%%%%%%%%%%%%%%%%%%%%%%%%%%%%%%%%%%%%%%%%%
\subsection{Estimation Bias from Ignoring Engagement}\label{sec:bias}
In this section we study the impact of
ignoring user engagement on the estimation bias of item utilities.
A key benefit of modeling user engagement is that it results in more accurate
estimation of item utilities by correcting for the way user engagement affects
choice decisions.  Ignoring user engagement will lead to biased estimates
of item utilities.  The amount of bias depends upon the value of the item utility
and the effect can be pronounced for very popular and unpopular items.

We consider two items $i$ and $j$ with utilities $w$ and $1-w$. We assume that there are $N$ users comparing the items,
and each user $u\in\curly{1,2,...,N}$ has engagement parameter $\epsilon_u$.  We assume that $\epsilon_u$ is drawn from
a  distribution with mean $\epsilon$ which  characterizes the engagement of the population.  Each user generates a single choice decision $C_u$  which is one if item $i$ is chosen and zero if item $j$ is chosen.  If the choice decisions are generated according to our choice engagement time model, then $C_u$ is a Bernoulli random variable which equals one with probability
\[p_u  = \frac{w+\epsilon_u}{1+2\epsilon_u}.\]  
If we neglect the user engagement, then this model reduces to the standard Bradley-Terry  model.

  We take a Bayesian approach
and put a uniform prior on $w$.  Our estimate of $w$ when we do not account for user engagement, which we denote $\widehat{w}$, will be the posterior mean.  For the standard Bradley-Terry model, the posterior distribution of $w$ is
a Beta distribution.  If we define the statistic $N_1 = \sum_{u=1}^NC_u$ then the resulting posterior mean is given by
\begin{align}
\widehat{w} = \frac{N_1+1}{N+2}\label{eq:what}.
\end{align}
We have the following result regarding the bias of the estimator, with proof in the appendix.
%%%%%%%%%%%%%%%%%%%%%%%%%%5
\begin{thm}\label{thm:bias}
Assume $N$ choice decisions $\curly{C_u}_{u=1}^N$ are generated according the choice engagement time model for two items with  utilities $w$ and $1-w$, respectively.  Let the user engagement parameters $\epsilon_u$ be i.i.d. non-negative random variables with 
mean $\epsilon$.  Let $\widehat{w}$ be the posterior mean of $w$ 
in the standard Bradley-Terry model given by equation \eqref{eq:what}. Define 
\[B= \mathbf E\bracket{\frac{\epsilon_u}{1+2\epsilon_u}}.\]  
Then  we have that 
\begin{align}
	\lim_{ N\rightarrow\infty}\mathbf E\bracket{\widehat{w}}-w & = B(1-2w),
\end{align}
where the expectation is taken with respect to the user engagement parameters and the choice data.
\end{thm}
This result shows that the bias of the standard Bradley-Terry estimate is determined by the value of $w$ and the distribution of $\epsilon_u$.
For $w=1/2$ the estimator is unbiased.  In this case choice behavior with and without engagement are the same because the choice probabilities equal $1/2$ for both models.  The choice data does not allow one to differentiate a choice decision
between two equally preferred items or a choice decision by a low engagement user.  For $w\neq 1/2$, the term
$B$ impacts the bias.  This value will be small if user engagement is  large (small $\epsilon_u$).  The sign of the bias
depends on $w$.  For $w<1/2$ the bias is positive, which means the Bradley-Terry estimate increases
the estimated utility of the item.  If $w>1/2$, then the opposite happens and the Bradley-Terry
estimate decreases.  Therefore, ignoring bias would increase the estimated utility of unpopular items
and decrease the utility of popular items.

\begin{figure}
\centering
\includegraphics[scale=.4]{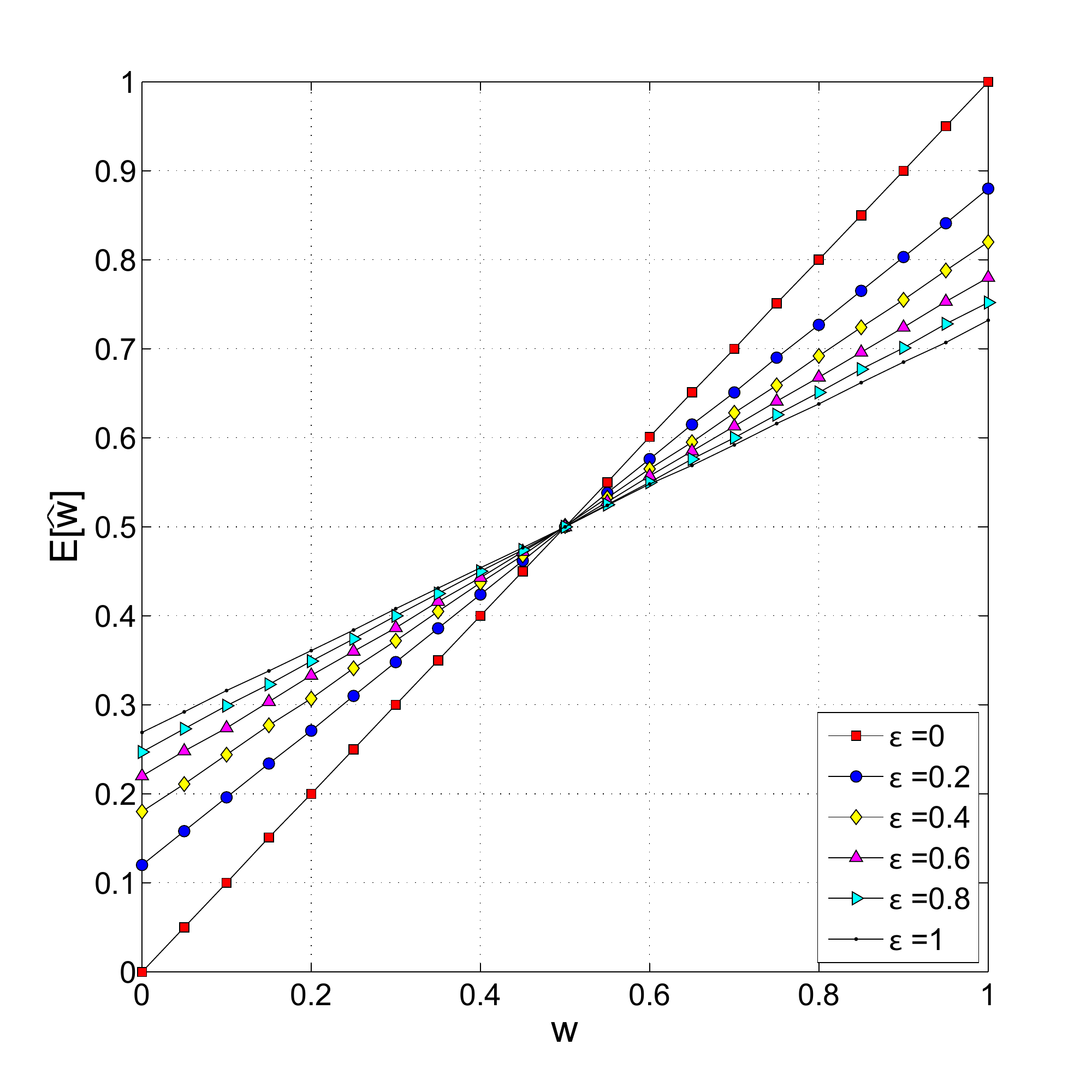}
\caption{Plot of $w$ versus $\mathbf E\bracket{\widehat{w}}$ for different
values of $\epsilon = \mathbf E \bracket{\epsilon_u}$.  The user engagement $\epsilon_u$ is drawn from an exponential distribution with means $\epsilon$ indicated on the figure.}
\label{fig:bias}
\end{figure}

 To further illustrate how user engagement impacts the bias, we plot in Figure \ref{fig:bias} the estimate of the Bradley-Terry model versus the true value for $w$ for different values of $\epsilon$.  As can be seen, for lower user engagement, the bias can be significant, making one think
that an item is popular when in reality it has a very low utility.

%%%%%%%%%%%%%%%%%%%%%%%%%%%%%%%%%%%%%%%%%%%%%%%%%%%%%%%%%%%%%%%%%%%%%
\section{Model Estimation}\label{sec:estimation}
We take a Bayesian approach to model estimation and calculate the posterior
distribution of the model parameters given the observed data.  We propose a hierarchical
structure for the choice engagement time model which allows for the sharing of information between
different users.  The Bayesian approach allows for simple estimation of hierarchical models. 
We now provide an overview of our model specification and  estimation procedure.  The technical details can be found
in the appendix.

%%%%%%%%%%%%%%%%%%%%%%%%%%%%%%%%%%%%%%%%%%%%%%%%%%%%%%%%%%%%%%
\subsection{Bayesian Model Specification}
\begin{figure}
\centering
\includegraphics[scale=0.75]{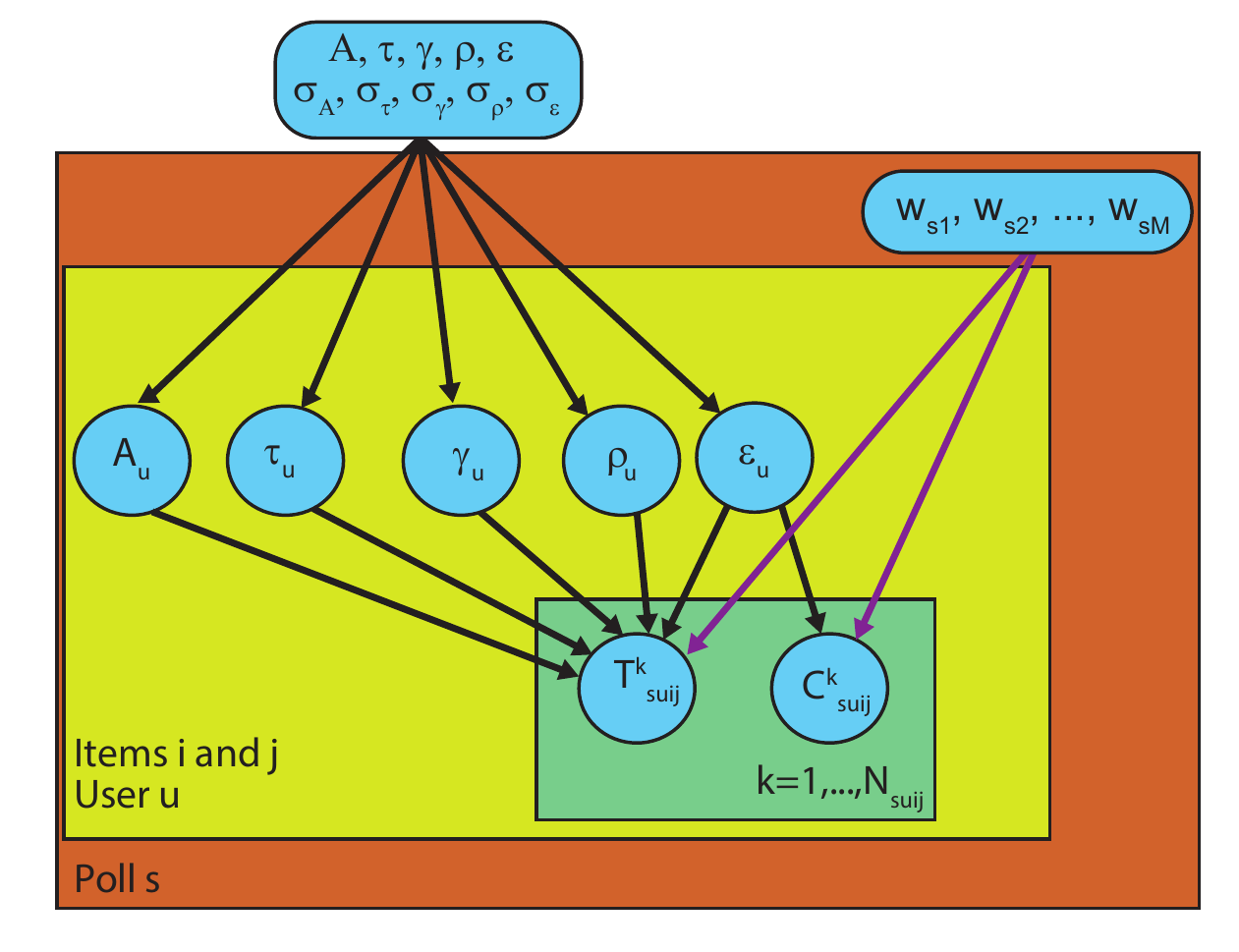}
\caption{Graphical model of choice engagement time model.  The plates denote replication over item pairs,
 users, and polls.  Hyperpriors are omitted for simplicity.}
\label{fig:graphical_model}
\end{figure}

Our empirical data presented in Section \ref{sec:data} consists of choice and time data from pairwise comparisons of items in
multiple polls from multiple users.  For the $k$th comparison between items $i$ and $j$ in poll $s$
by user $u$, we denote the choice decision and response time as $C_{suij}^k$ and $T_{suij}^k$.
For the model parameters we impose the following structure.  Item utilities are common to all users and are independent a priori.
We denote the utility of item $i$ in poll $s$ as $w_{si}$.   The engagement and timing parameters are user specific, and we expect there to be some level of homogeneity in the timing behavior across users.  For instance, response times
on a mobile application may have a natural time scale of a few seconds for all users.  To capture
this shared behavior, we impose a hierarchy on the user parameters.  We assume that each user parameter is drawn from a global distribution which characterizes the population.  This induces a correlation among the user parameters a priori. 
Our model then contains user parameters which model the time and choice data generated by the users
and global parameters which model the population level behavior.  To specify our model we use
the following notation.  We let $\mathcal N(\mu,\sigma^2)$ represent a normal distribution with mean $\mu$ and
variance $\sigma^2$, $\text{Hypo}(a,b)$ represent a hypoexponential distribution with parameters $a$ and $b$ and mean $a+b$ (see equation \eqref{eq:hypoexponential}), and $\text{Bern}(p)$
represent a Bernoulli distribution with mean $p$.
With this notation,  our choice engagement time model specification for a user $u$ viewing items $i$ and $j$ in poll $s$ is
\begin{align}
	C^k_{suij}|p_{suij} &\sim \text{Bern}(p_{suij}),\nonumber \\
	T^k_{suij}|\tau_{u},\mu_{suij} &\sim \text{Hypo}(\tau_{u},\mu_{suij}),\nonumber \\
	p_{suij} & = \frac{\frac{w_{si}}{w_{si}+w_{sj}}+\epsilon_u}{1+2\epsilon_u},\label{eq:psuij}\\	
	\mu_{suij} & = \frac{A_u}{\paranth{\frac{w_{si}-w_{sj}}{w_{si}+w_{sj}}}^{2\gamma_u}+\epsilon_u+\rho_u}\label{eq:musuij}.
\end{align}
The user parameters are given normal priors whose means and variances are given
by global parameters.  The priors are
\begin{align*}
	A_u & \sim\mathcal N(A,\sigma^2_A),\\
	\tau_{u} &\sim\mathcal N(\tau,\sigma^2_\tau),\\
	\rho_u & \sim\mathcal N(\rho,\sigma^2_\rho),\\
	\epsilon_u & \sim\mathcal N(\epsilon,\sigma^2_\epsilon),\\
	\gamma_u & \sim\mathcal N(\gamma,\sigma^2_\gamma).
\end{align*}
%%%
The means and variances of the user parameter priors characterize the population level behavior,
but still allow for flexibility to fit individual user behavior.
 We illustrate the structure of our choice engagement time model in the graphical model in Figure \ref{fig:graphical_model}.     

%%%%%%%%%%%%%%%%%%%%%%%%%%%%%%%%%%%%%%%%%%%%%%%%%%%%%%%%%%%%%%%%%%%%%
\subsection{Posterior Distribution}\label{sec:procedure}
Model estimation entails calculating the posterior distribution of the
model parameters conditioned on the data.
We now set up notation that will allow us to derive the form of the posterior distribution.
We assume that the users are shown two items at a time.  There are
 $S$ polls, and each poll $s$ has $M_s$ items. $N$ users complete all polls, and there are $N_{suij}$ comparisons of items
$i$ and $j$ in poll $s$ by user $u$.  In practice this value would be one, as a user would 
compare a pair of items once, but we allow for arbitrary values here.  The model's
global parameters  are the item utilities $\mathbf w = \curly{w_{si}}$ ($1\leq s \leq S$, $1\leq i\leq M_s$) and the timing
parameters $\Theta = \curly{A,\tau,\rho,\epsilon,  \gamma, \sigma^2_A,\sigma^2_\tau,\sigma^2_\rho,\sigma^2_\epsilon,\sigma^2_\gamma}$. To ensure that the item utilities of each poll sum to one we set $w_{sM_s} = 1-\sum_{i=1}^{M_s-1}w_{si}$ for each poll $s$. 
 The  parameters for a user $u$ are $\Phi_u=\curly{A_u,\tau_u,\rho_u,\epsilon_u,\gamma_u}$ and we define the set of all user
parameters as $\Phi = \bigcup_{i=1}^N\Phi_u$.  The observed choice and time data are 
$\mathbf C = \bigcup_{s=1}^S\bigcup_{u=1}^N \bigcup_{i,j=1}^{M_s}\bigcup_{k=1}^{N_{suij}} C_{suij}^k$
and $\mathbf T = \bigcup_{s=1}^S\bigcup_{u=1}^N \bigcup_{i,j=1}^{M_s}\bigcup_{k=1}^{N_{suij}} T_{suij}^k$,
respectively.

We start with the data likelihood.  Conditioned on the user parameters, the choice and time data are independent.  The choice data are Bernoulli
random variables with different means and their likelihood is 
\begin{align}
	\mathbf P\paranth{\mathbf C|\Phi,\mathbf w} & = \prod_{s=1}^{S} \prod_{u=1}^{N} \prod_{i,j=1}^{M_s}
	\prod_{k=1}^{N_{suij}} p_{suij}^{C^k_{suij}}\paranth{1-p_{suij}}^{1-C^k_{suij}}\label{eq:likelihood_C}
\end{align}
where $p_{suij}$ is given by equation \eqref{eq:psuij}.  As shown earlier, the response times are hypoexponential random variables.
We let $g(t;a,b)$ be the probability density function of a hypoexponential random
variable with parameters $a$ and $b$ given in equation \eqref{eq:hypoexponential}.   This allows us to write the likelihood of the time data 
as
\begin{align}
	\mathbf P\paranth{\mathbf T|\Phi,\mathbf w} & =
	\prod_{s=1}^{S} \prod_{u=1}^{N} \prod_{i,j=1}^{M_s}
	\prod_{k=1}^{N_{suij}}
	g(T^k_{suij};\tau_{u},\mu_{suij})\label{eq:likelihood_T}
\end{align}
with $\mu_{suij}$ given by equation \eqref{eq:musuij}.

The user parameters are all normally distributed, conditioned on the global parameters.
Define the normal density with mean $\mu$ and variance $\sigma^2$ as
\begin{align}
	f(t;\mu,\sigma^2) & = \frac{1}{\sqrt{2\pi\sigma^2}}e^{-\frac{(t-\mu)^2}{2\sigma^2}}.
\end{align}
To simplify notation for the  likelihood of the user parameters, we define the set
$\Xi= \curly{A,\tau,\rho,\epsilon,\gamma}$.  The the likelihood can be written as
\begin{align}
	\mathbf P\paranth{\Phi|\Theta} & = \prod_{\alpha\in\Xi}\prod_{u=1}^{N} f(\alpha_u;\alpha,\sigma^2_\alpha)
\end{align}
%%%
To complete our Bayesian specification, we need to put hyperpriors
on the global parameters.  Following
a standard Bayesian approach, we will choose
uninformative hyperpriors for each parameter.  We choose conjugate hyperpriors
if possible to simplify the calculation of the posterior distribution.  
Each item utility hyperprior is a uniform distribution on $[0,1]$.  
  For the global means $(A,\tau,\rho,\epsilon,\gamma)$
we choose normal hyperpriors with mean zero and standard deviation 100.    For the global
variances $(\sigma^2_A,\sigma^2_\tau,\sigma^2_\rho,\sigma^2_\epsilon,\sigma^2_\gamma)$, we choose inverse-gamma priors with shape and scale equal to one. The inverse-gamma density with shape and scale $a$ and $b$ is given by
\begin{align}
	h(t;a,b) = \frac{b^a}{\Gamma(a)}t^{-(a+1)}e^{-\frac{b}{t}}
\end{align}
where $\Gamma(t)$ is the gamma function. 
 We  denote the hyperprior distribution of the global timing parameters as $\mathbf P(\Theta)$ and
of the item utilities as $\mathbf P(\mathbf w)$.  These distributions are given by
\begin{align}
	\mathbf P(\Theta,\mathbf w) & = \prod_{\alpha\in\Xi} f(\alpha;0,100^2)h(\sigma^2_\alpha;1,1).
\end{align}

The posterior distribution for the model parameters is obtained by combining
all the likelihood functions and applying Bayes rule.  The resulting
posterior distribution is
\begin{align}
	\mathbf P\paranth{\Phi,\Theta,\mathbf w|\mathbf C,\mathbf T} & = 
	\frac{\mathbf P\paranth{\mathbf C|\Phi,\mathbf w}
	       \mathbf P\paranth{\mathbf T|\Phi,\mathbf w}
	       \mathbf P\paranth{\Phi|\Theta}\mathbf P\paranth{\Theta,\mathbf w}}
			{\mathbf P\paranth{\mathbf C,\mathbf T}}.
\end{align}
To evaluate the posterior distribution we generate samples form it using a Metropolis within Gibbs' algorithm.  The details of our sampling algorithm can be found in the appendix.
%%%%%%%%%%%%%%%%%%%%%%%%%%%%%%%%%%%%%%%%%%%%%%%%%%%%%%%%%%%%%%%%%%%%%
\section{Results}\label{sec:results}
We next examine the results of our Bayesian model estimation on the observed choice and time data for the
student and AMT populations.  
 We estimate the models separately for the student and AMT  data in order to present
the distinct behaviors of these two populations.  For each population, we generated posterior samples using three
independent MCMC chains with dispersed starting points run for 5000 
iterations and discarded a burn-in period of 500 iterations. Convergence of
the MCMC sampler was assessed using trace plots. 

%%%%%%%%%%%%%%%%%%%%%%%%%%%%%%%%%%%%%%%%%%%%%%%%
\subsection{Student Versus AMT Subjects}

The posterior means of the global 
model parameters for each population are shown in Table \ref{table:posterior}.
Just as we found in Section \ref{sec:data}, there is a clear difference
between the student and AMT populations in terms of their timing and engagement
behavior.  The students have a smaller value for $\epsilon$ than the AMT population
(0.20 versus 0.27) and a larger value for $\gamma$ (1.04 versus 0.70).  This indicates
that the students are more engaged with their choice decisions and their response times
are more sensitive to the choice decision.  In fact, the difference in $\gamma$ is quite larger,
indicating that the AMT workers' response times do not vary much as a function of the items shown.
This may be due to the fact that they are online workers and may not be very diligent when making
their decisions.  

%To gain a more concrete measure of the difference
%in the populations, we can calculate the minimum and maximum mean response time for the two populations
%using equation \eqref{eq:muT} and the posterior means for the global parameter values.  Doing so we find
%that for the students, the mean response time ranges from 2.49 to 3.10 seconds, while for the AMT workers
%the range is 2.00 to 2.49.  For comparison, from Table \ref{table:ug_vs_mt} we saw that the mean response times
%were 3.18 and 2.11 seconds for the student and AMT populations, respectively.  Thus, the estimated parameter
%values align with our exploratory analysis.

\begin{figure}
\centering
\includegraphics[scale=0.5]{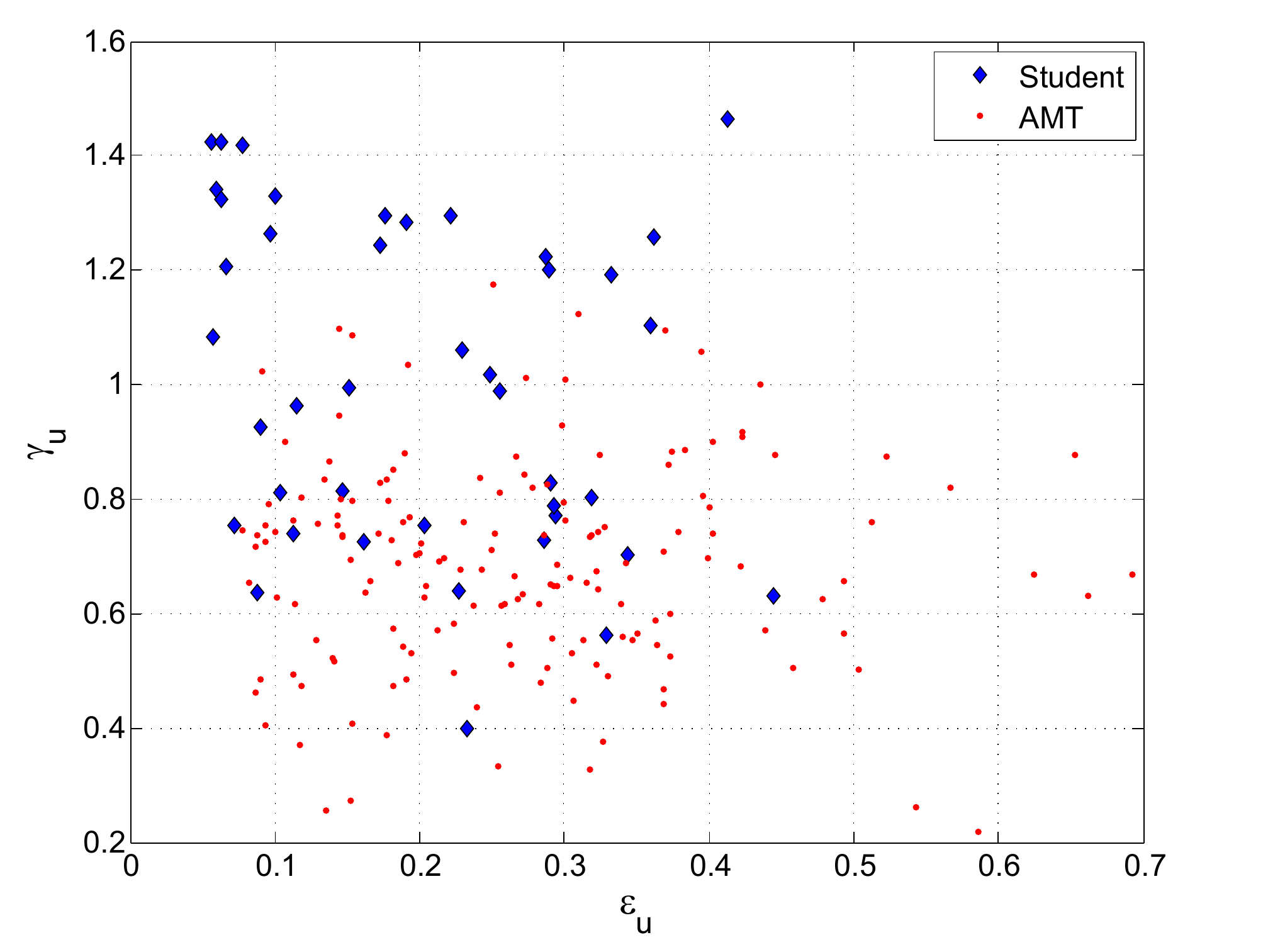}
\caption{Plot of the estimated $\epsilon_u$ and $\gamma_u$ parameters for the student and AMT populations.}
\label{fig:user_params}
\end{figure}

The estimated individual user parameters provide us more insights about the engagement of the users.
We plot in Figure \ref{fig:user_params} the estimates of $\epsilon_u$ and $\gamma_u$ for each user
in the two subject populations.  It can be seen that each population occupies different regions of this parameter space
which correspond to their different behavioral characteristics.  
The student population has many users with a small $\epsilon_u$ and  a large
$\gamma_u$, whereas the AMT population has several users with a large $\epsilon_u$ and a smaller $\gamma_u$.
There are also some users in the two populations which have similar parameter values.  These users
appear to have similar levels of engagement, despite being in different populations.  However,
overall it appears that these two subject groups exhibit different levels of engagement.

\begin{figure}
\centering
\includegraphics[scale=0.5]{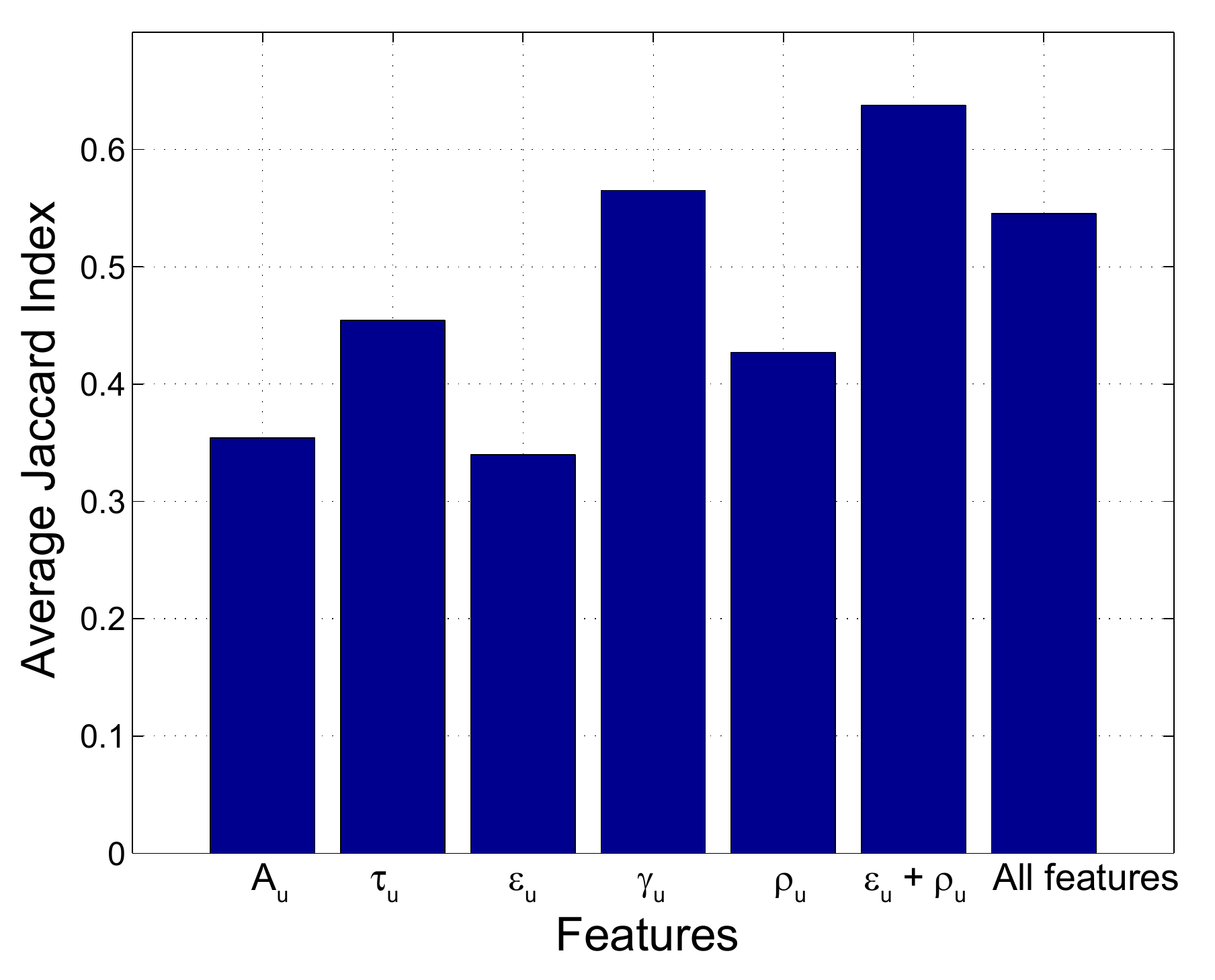}
\caption{Bar graph of the average Jaccard index for $k$-means clustering of users into student and AMT groups using
 different user parameters as features.  A higher average Jaccard index means a more accurate clustering.}
\label{fig:jaccard}
\end{figure}

We saw earlier that the AMT subjects seemed to have lower engagement
than the students.  We now show that we can recover this classification
using the user parameter values estimated for the choice engagement time model.
We apply $k$-means clustering with two clusters to the user parameter data.
We first use as features an individual parameter, such as 
$A_u,\tau_u,\epsilon_u, \gamma_u$, or $\rho_u$.  We then try a combination
of $\epsilon_u$ and $\gamma_u$, the features plotted in Figure \ref{fig:user_params}.  
Finally, we include all user parameters
as features.  To measure the accuracy of the classification we use the
average Jaccard index which is defined as follows.  For two set $B$ and $C$,
their Jaccard index is defined as $J(B,C) = |B\bigcap C|/|A\bigcup C|$.  This value
is larger for more similar sets.  We let
$L_{1}$ and $L_{2}$ denote the sets of student and AMT subjects and we let
$K_1$ and $K_2$ denote the two clusters produced by the $k$-means algorithm.  The average
Jaccard distance is then
\begin{align}
	   J_1(L_{1},L_{2},K_1,K_2) & = \frac{1}{2}\max\curly{J(L_{1},K_1)+J(L_{2},K_2),J(L_{1},K_2)+J(L_{2},K_1)}.\nonumber
\end{align}
  This measure
has a maximum value of one when the sets of clusters are equal.  We plot the average Jaccard index for the different
feature sets in Figure \ref{fig:jaccard}.  We see that the single feature that does best is $\gamma_u$, while $\epsilon_u$ combined
with $\gamma_u$ gives the overall best accuracy.  Interestingly, when all features are used the accuracy decreases.
This analysis shows the importance of $\gamma_u$ for measuring engagement.  While $\epsilon_u$ gives the influence
of engagement on choice decisions, by itself it cannot separate low and high engagement users as effectively as
$\gamma_u$.  This may be because more engaged users will have higher variation in their response times 
if they have variable preferences for the items shown.  The sensitivity of the response time to item preferences
appears to an effective way to segment users by engagement level.   While $\epsilon_u$ is important for correcting biases
in estimating preferences, as we will see in Section \ref{sec:results_utilities}, $\gamma_u$
is useful for segmenting users by engagement level.  Since $\gamma_u$ can only
be estimated from response time data, this shows the importance of response time data
in choice modeling.

\begin{table}
\centering
\begin{tabulary}{1.0\textwidth}{|C|C|C|C|}
\hline
 Parameter & Student posterior mean (c.i.) &  AMT  posterior mean (c.i.)\\
\hline
$A$ [sec] & 1.37 (0.88, 1.71) & 1.18 (0.81, 1.32)\\
\hline
$\tau$ [sec] & 0.85 (0.64, 1.01) & 0.98 (0.89, 1.03)\\
\hline
$\rho$ & 0.41 (0.25, 0.58) & 0.51 (0.33, 0.67)\\
\hline
$\epsilon$ & 0.20 (0.11, 0.29) & 0.27 (0.20, 0.33)\\
\hline 
$\gamma$ & 1.04 (0.71, 1.21) & 0.70 (0.56, 0.79)\\
\hline
$\sigma_A$ & 0.40 (0.30, 0.55) & 0.34 (0.27, 0.42)\\
\hline
$\sigma_\tau$ & 0.37 (0.29, 0.47) &0.30 (0.26,0.34)\\
\hline
$\sigma_\rho$ & 0.35 (0.27, 0.47) &0.25 (0.20, 0.30)\\
\hline
$\sigma_\epsilon$ &0.28 (0.23, 0.34)&	0.22 (0.18,0.26)\\
\hline
$\sigma_\gamma$ & 0.47 (0.37, 0.64)&	0.37 (0.29,0.42)\\
\hline
\end{tabulary} 
\caption{Posterior means and 90\% credibility intervals
(c.i.) for the global model parameters}\label{table:posterior}
\end{table}

%%%%%%%%%%%%%%%%%%%%%%%%%%%%%%%%%%%%
\subsection{Comparison with Other Models}\label{sec:comparison}
We next compare our model to two different benchmark models.
The first model is a simple Bradley-Terry choice model where user engagement is set to zero
and response times are hypoexponentially distributed with parameters $\tau_u$ and $\mu_u$ for user $u$.  We refer to
this simply as a choice model.  
The second model, which we refer to as a choice engagement model, incorporates user engagement into the choice probability, but the response time has the same distribution as the basic choice model.  We will refer to
our full model as a choice engagement time model.    We summarize the
choice probability and mean response time for the different models in Table \ref{table:models}.  
To assess model fit,
we use the deviance information criterion (DIC) \citep{ref:dic}.  This is a model fit
score that favors models which fit the data well but penalizes models with many degrees of freedom.  A DIC that is smaller signifies a superior model fit.  We show the DIC of the three models considered in Table \ref{table:models}.  We first see that
including user engagement in the choice probability improves the model fit.  Second, allowing the response
times to depend upon the item utilities further improves the model fit.  This shows that engagement
and item dependent response times are important elements when modeling choice.

%%%%%%%%%%%%%%%%%%%%%%%%%%%%%%%%%%%%
\begin{table}
\centering
\begin{tabulary}{1.0\textwidth}{|L|C|C|C|C|C|}
\hline
 Model & Choice probability  &  Mean response time & Student DIC & AMT DIC \\
\hline
Choice    & \[\frac{w_{si}}{w_{si}+w_{sj}}\] & \[\tau_u+\mu_u\]& 4,536 & 30,123\\
\hline
Choice engagement & Equation \eqref{eq:p} & \[\tau_u+\mu_u\]& 4,532 & 29,952\\
\hline
Choice engagement time  & Equation \eqref{eq:p} & Equation \eqref{eq:muT}& 4,435 & 29,941\\
\hline
\end{tabulary} 
\caption{The choice probability and mean response time for the different models and the corresponding deviance
information criterion (DIC) for the student and AMT populations.  The parameter values are for data for poll $s$, items $i$ and $j$, and user $u$.}\label{table:models}
\end{table}
%%%%%%%%%%%%%%%%%%%%%%%%%%%%%%%%%%%

%%%%%%%%%%%%%%%%%%%%%%%%%%%%%%%%%%%%%%%%%%%%%%%%%%%%%%%%%%%%%%%%%%%%%%%%%%%%

\subsection{Item Utilities}\label{sec:results_utilities}
\begin{figure}
\centering
\includegraphics[scale=0.5]{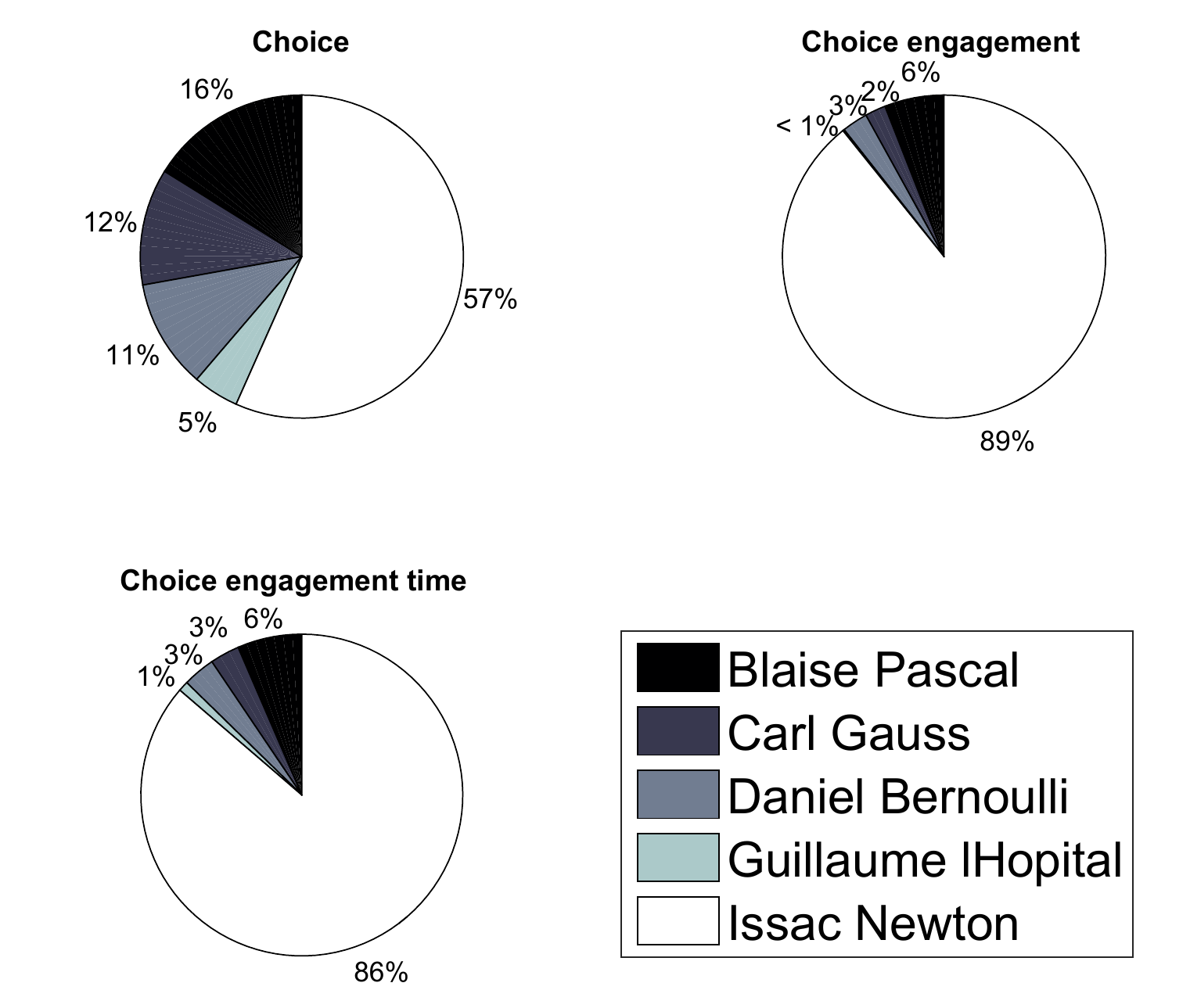}
\caption{Pie graphs of the posterior median of item utilities for different models for the Mathematicians poll on the AMT population.}
\label{fig:poll_cet}
\end{figure}

We next look at the difference in the item utilities for the different models in each population.
If we do not model engagement or response times, the choice data can seem more random and items can appear
to have similar utilities.   When engagement and response times are included,  we expect the item utilities to become more spread out and less equal.  We show an example of this for the Mathematicians poll for the AMT population in Figure \ref{fig:poll_cet} where we plot the posterior median of the item utilities estimated from the different models.  It can be seen here that when engagement is included, the utility for the most popular item increases dramatically.  Without modeling engagement, one would not realize
the extent to which the most popular item dominates all other items.  This suggests that by modeling engagement in the choice decision, we can gain a more accurate picture of user preferences.

\begin{figure}
\centering
\includegraphics[scale=.5]{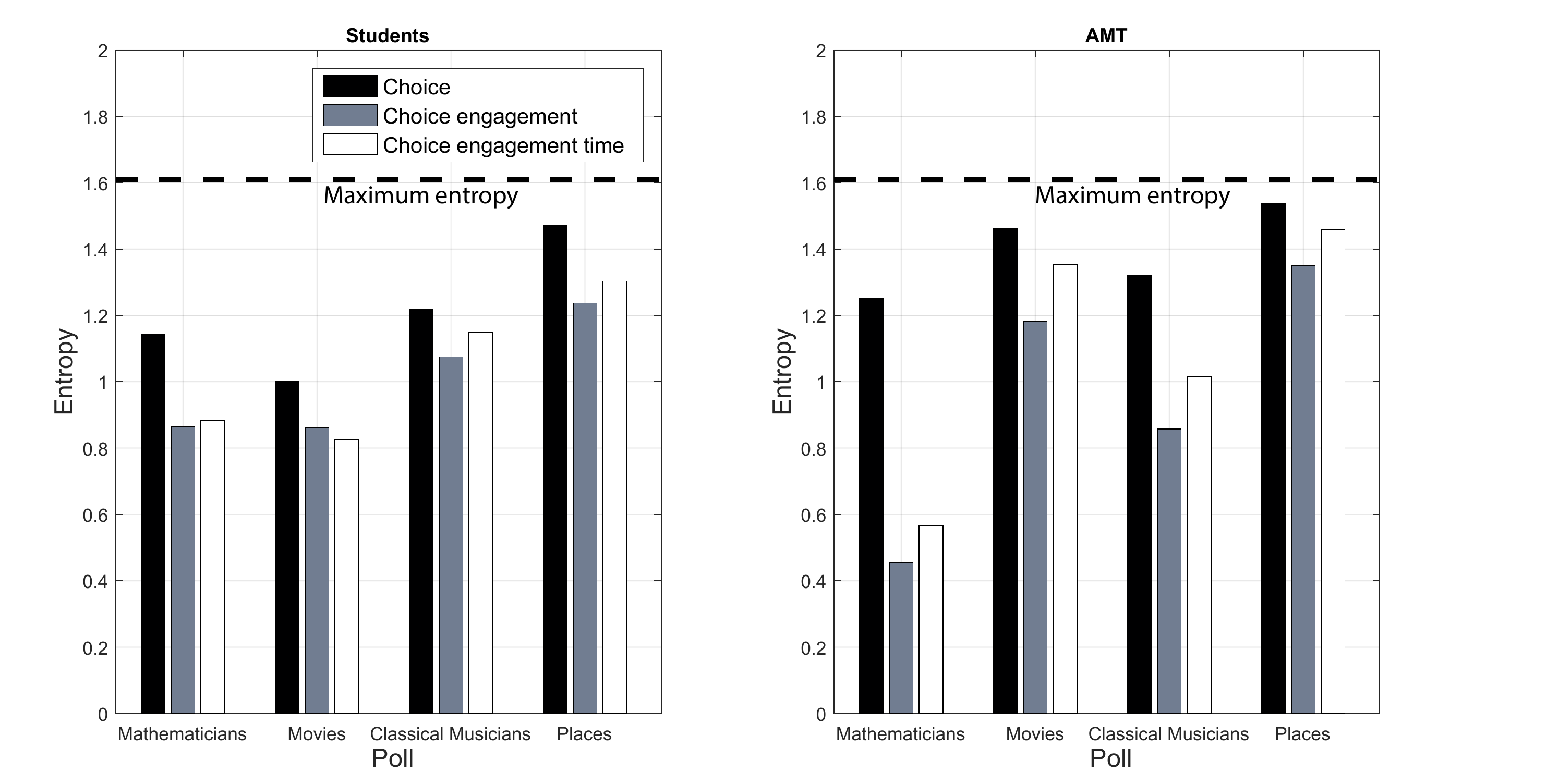}
\caption{Plot of the entropy of each poll for the choice, choice engagement, and choice engagement time models on the (left) student and (right) AMT populations.  The maximum possible entropy is indicated by the dashed line. }
\label{fig:entropy_model}
\end{figure}

The divergence of the item utilities from the engagement biased values can be quantified by using the entropy measure.
For a poll of $M$ items, let the utility vector be defined as $\mathbf w =\curly{w_{i}}_{i=1}^M$ with $\sum_{i=1}^Mw_{i}=1$.
Defined this way, the item utilities form a discrete distribution over the poll items.  The entropy of this distribution is
\begin{align}
	H(\mathbf w) & = -\sum_{i=1}^M w_{i}\log(w_{i})\label{eq:entropy}.
\end{align}
The entropy achieves its maximum value of $\log(M)$ when all item utilities are all equal and achieves its minimum value of zero when one item has utility equal to one and all others have utility zero.  High entropy corresponds to more random choices, while lower entropy corresponds to one item or a small set of items dominating the preferences.
Figure \ref{fig:entropy_model} shows the entropy of each poll for each model and population.  As can be seen, the choice model which does not model engagement always has the highest entropy. The models with engagement produce item utilities which are less uniform, resulting in a lower entropy.  This effect is most dramatic for the AMT population on the Mathematicians poll, where we saw
one item become completely dominant once engagement is included.  The inclusion of time data slightly raises the entropy with respect to the choice engagement model for most polls, although for the student population on the Movies poll the inclusion of time data slightly reduces the entropy.  

\begin{figure}

\centering
\includegraphics[scale=.51]{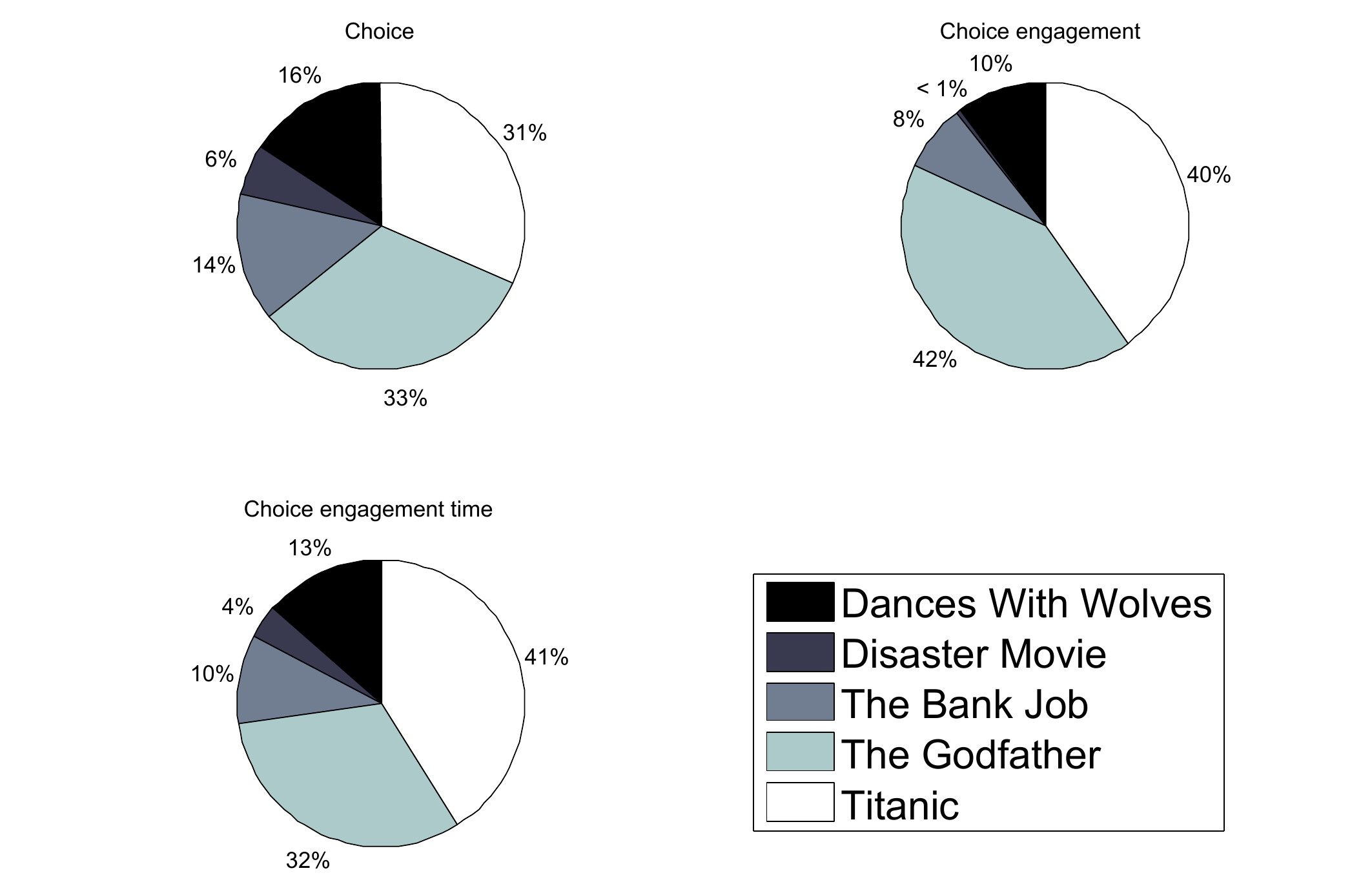}
\caption{Pie graphs of the posterior median of item utilities for different models for the Movies poll on the AMT population. }
\label{fig:movies}
\end{figure}

We also find that in some instances time data can also affect the resulting item utilities in significant ways.  For the majority of our polls time data
did not change the ranking of the items based on their estimated utilities.  
However, there was one poll where the inclusion of time data resulted in a change.
The Movies poll for the AMT population had its top two items switched when time data was included.  We show the different
model rankings in the pie charts in Figure \ref{fig:movies}.  As can be seen, for the choice and choice engagement models,
The Godfather is the top ranked item, though only by a small amount.  When time data is included in the choice engagement time model,
Titanic becomes the top ranked movie, and by a wider margin.  We suspect that the time data is able to discern the stronger
preference of the users for Titanic, causing the resulting changing in the item ranking.

%%%%%%%%%%%%%%%%%%%%%%%%%%%%%%%%%%%%%%%%%%%%%%%%%%%%%%%%%%%%%
%%%%%%%%%%%%%%%%%%%%%%%%%%%%%%%%%%%%%%%%%%%%%%%%%%%%%%%%%%%%%%%%%%%%%%%%
%%%%%%%%%%%%%%%%%%%%%%%%%%%%%%%%%%%%%%%%%%%%%%%%%%%%%%%%%%%%%%%%%%%%%%%%%%%%%%%%%%%%%%%%%%%%%%%%%%%%%%%%%%%%%%%%%%%%%%%%%%%%%%%%%%%%%%%%%%%%%%%%
\section{Conclusion}\label{sec:conclusion}
We have found that modeling
engagement and response times in choice decisions 
can provide new insights about users' behavior and preferences.  Low engagement produces
biases in choice decisions and faster response times.  We saw
that in certain online polls, low engagement produces random choice decisions.  
We also saw that there was a correlation between choice decisions and response times.
We proposed a Bayesian hierarchical choice engagement time model these phenomena.  
Our model drew upon properties of previous psychological models,
but allowed for simple estimation and interpretation.  We saw that including engagement
and response times in our model resulted in a better fit to observed online poll
data.  In addition, our choice engagement time model was able to correct
for the effects of low engagement in estimating user preferences and also
segment users by engagement level using estimated parameter values.

Modeling engagement and response times with choice decisions has two practical benefits.
First, it provides more accurate measurements of user preferences.  By not modeling engagement,
one may be led to believe that certain items are equally preferred by users.  However, by accounting
for user engagement and response times we obtain much more asymmetric preferences which may better
align with true user preferences.
Second, modeling engagement and response times  allows
us to characterize the engagement of users
at the individual level in a simple and interpretable manner.  This can be of use especially to online
applications which want to obtain deeper insights about their user base.    For instance,  users that have
small $\epsilon_u$ and large $\gamma_u$ are more engaged users who are most likely providing
genuine preference data.  These types of behavioral insights  
give online applications a more complete understanding about the true level of engagement of their user base.
 These insights can also help inform many operational decisions
of online applications, such as choosing assortments to show users or
 identifying target users for marketing or promotional campaigns.  Because time data
is already measured in  online applications, choice engagement time models
can be a powerful and useful framework which can be easily implemented and utilized.

%%%%%%%%%%%%%%%%%%%%%%%%%%%%%%%%%%%%%%%%%%%%%%%%%%%%%%%%%%%%%%%%%%%%%%%%%%%%%%%%%%%%%5
\newpage
\section{Appendix: Proof of Theorem \ref{thm:poisson}}
We begin by calculating $p$, the probability that item $a$ is chosen.
This is the probability that $N_a(t)$ has $K_a$ arrivals before $N_b(t)$ has
$K_b$ arrivals, averaged over all $t$.  Another way to view this is to consider
the merged process $N(t) = N_a(t)+N_b(t)$ which is a Poisson process with rate $\alpha+\beta$.
Standard results for merged Poisson processes shows that 
the probability of an arrival to the merged process being of type $a$ is $\alpha/(\alpha+\beta)$.
We want to have less than $K_b$ arrivals when we get $K_a$ arrivals.  Let $B$ be the number of
$b$ arrivals before $K_a$ arrivals.  Then $B$ is a negative binomial random variable with parameters $K_a$ and $\beta/(\alpha+\beta)$.
The CDF of a negative binomial random variable with parameters $(K,x)$ evaluated at $k$ is $1-I_x(k+1,K)$, where
we have used the notation for the regularized incomplete beta function from equation \eqref{eq:beta_inc}.
The probability of choosing $a$ is equal to the probability that $B$ is less than $K_b$.  Using the expression for the negative binomial CDF we obtain
\begin{align*}
	p & = \mathbf P(B\leq K_b-1)\\
	  & = 1-I_{\beta/(\alpha+\beta)}(K_b,K_a)\\
		& = I_{\alpha/(\alpha+\beta)}(K_a,K_b).
\end{align*}
Above we used the property $I_x(u,v) = 1-I_{1-x}(v,u)$.

The conditional mean response time given $a$ can be found by using the expression for the likelihood of the choice and
time data given by equation \eqref{eq:likelihood_poisson}.  This density is defined for $t>0$ and $c\in\curly{a,b}$.  The mean response time conditional on $c=a$ is 
\begin{align}
	\mu_a & = \frac{\int_0^\infty tf(a,t)dt}{\int_0^\infty f(a,t)dt}=\frac{1}{p}\int_0^\infty tf(a,t)dt\label{eq:mua1}.
\end{align}
Above we have used the fact that $p = \int_0^\infty f(a,t)dt$.  To evaluate the integral in the above
expression we make use of the following result which can be proven using elementary properties of the moment generating
function of an Erlang random variable.
%%%
\begin{lem}\label{lem:erlang}
Let $X$ be an Erlang random variable with parameters $K$ and $\alpha$.  Then
\begin{align}
	\mathbf E\bracket{X^ne^{-\beta X}}& = \frac{(K+n-1)!\alpha^K}{(K-1)!(\alpha+\beta)^{K+n}}.
\end{align}
\end{lem}
%%%
We define $T_a$ to be an Erlang random variable with parameters $K_a$ and $\alpha$. Using
equations \eqref{eq:erlang_pdf} and \eqref{eq:erlang_cdf} we have
\begin{align}
	\int_0^\infty tf(a,t)dt& = \int_0^\infty tf_E(t;K_a,\alpha)(1-F_E(t;K_b,\beta)dt\nonumber\\
	& =  \int_0^\infty f_E(t;K_a,\alpha)\sum_{n=0}^{K_b-1}\frac{1}{n!}\beta^nt^{n+1}e^{-\beta t}\nonumber\\
	& = \sum_{n=0}^{K_b-1}\frac{1}{n!}\beta^n\mathbf E\bracket{T_a^{n+1}e^{-\beta T_a}}\nonumber\\
	& = \sum_{n=0}^{K_b-1} \beta^n\frac{(K_a+n)!}{(K_a-1)!n!}\paranth{\frac{1}{\alpha+\beta}}^{n+1}\paranth{\frac{\alpha}{\alpha+\beta}}^{K_a}\nonumber\\
	& = \frac{K_a}{\alpha}\sum_{n=0}^{K_b-1} {K_a+n\choose n}\paranth{\frac{\beta}{\alpha+\beta}}^n\paranth{\frac{\alpha}{\alpha+\beta}}^{K_a+1}\label{eq:erlang2}.
\end{align}
The expectation is evaluated using Lemma \ref{lem:erlang}.  The expression in the summation in equation \eqref{eq:erlang2} 
is the probability mass function  of a negative binomial random variable with parameters $K_a+1$ and $\beta/(\alpha+\beta)$.
Therefore, the sum is the probability that this random variable 
is less than or equal to $K_b-1$.  Using the expression for the negative binomial CDF and elementary properties
of the regularized incomplete beta function, we have
\begin{align}
	\int_0^\infty tf(a,t)dt& =\frac{K_a}{\alpha}\paranth{1-I_{\beta/(\alpha+\beta)}(K_b,K_a+1)}\nonumber\\
	& = \frac{K_a}{\alpha}I_{\alpha/(\alpha+\beta)}(K_a+1,K_b)\nonumber.
\end{align}
Inserting this into equation \eqref{eq:mua1} we obtain our result for $\mu_a$.  To obtain $\mu_b$, we repeat the
analysis with $\alpha$ and $\beta$, $K_a$ and $K_b$, and $p$ and $1-p$ all interchanged.
The unconditional mean is then given by integrating equation \eqref{eq:likelihood_poisson} over $t$
and summing over $c$, giving
\begin{align*}
	\mu & = \int_0^\infty tf(a,t)dt+\int_0^\infty tf(b,t)dt\\
	& =  \frac{K_a}{\alpha}I_{\alpha/(\alpha+\beta)}(K_a+1,K_b)+ \frac{K_b}{\beta}I_{\beta/(\alpha+\beta)}(K_b+1,K_a)
\end{align*}

%%%%%%%%%%%%%%%%%%%%%%%%%%%%%%%%%%%%%%%%%%%%%%%%%%%%%%%%%%%%%%%%%%%%%%%%%%%%%%%%%%%%%%

\section{Appendix: Proof of Theorem \ref{thm:ddm}}
Let $X(t)$ be a Brownian motion motion with drift $d$ and standard deviation $\sigma^2$ with $X(0) = z$.
The probability $p$ of choosing $a$ is equal to the probability that $X(t)$ hits $K>z$ before
it hits 0.  This can be evaluated using standard results from martingale theory.  
Define $B(t)$ as a standard Brownian motion.  Then $X(t) = \sigma B(t)+dt+z$.
We define the stopping time
$T=\inf\curly{t>0: X(t)=0 ~~\text{or} ~~X(t) = K}$.  From standard results for Brownian
motion we have that for any $\theta$, $M(t) = e^{\theta B(t)-\theta^2t/2}$ is a martingale.
We have by the optional stopping theorem that $\mathbf E\bracket{M(T)}= M(0) = 1$.
Substituting in $X(t)$  gives
\begin{align}
  \mathbf E\bracket {e^{\theta (X(T)-dT-z)/\sigma-\theta^2T/2}} &  =1\nonumber.  
\end{align}
Setting $\theta = -2d/\sigma$ we have
\begin{align}
  \mathbf E\bracket {e^{-2dX(T)/\sigma^2+2dz/\sigma^2}} &  =1\nonumber.  
\end{align}
We then use the fact that $X(T)=K$ with probability $p$ and $X(T)=0$
with probability $1-p$ to get
\begin{align}
	1-p+pe^{-2da/\sigma^2}& = e^{-2dz/\sigma^2}\nonumber
\end{align}
or by rearranging
\begin{align}
	p& = \frac{1-e^{-2dz/\sigma^2}}{1-e^{-2dK/\sigma^2}}.\nonumber
\end{align}

The unconditional mean $\mu$ can be found also using martingale theory.
Using our notation, $\mu = \mathbf E\bracket{T}$.
We use the fact that $B(t)$ is a martingale and the optional stopping theorem
to have $E\bracket{B(T)}  = 0$.  Substituting in $X(t)$ we get
\begin{align*}
	\mathbf E\bracket{ X(T)-dT-z} & = 0\\
	pK-d\mu - z & =0\\
	\mu & = \frac{pK-z}{d}.
\end{align*}

Similar to what was done for the Poisson counter model,
$\mu_a$ and $\mu_b$ are found by using equation \eqref{eq:mua1} and the expression for the likelihood of the choice and
time data given by equation \eqref{eq:likelihood_ddm}.  The integral in equation \eqref{eq:mua1} can be performed
using standard methods, giving the desired result.

%%%%%%%%%%%%%%%%%%%%%%%%%%%%%%%%%%%%%%%%%%%%%%%%%%%%%%%%%%%%%%%%
\section{Proof of Lemma \ref{lem:ddm_equal}}
We set $K = 2z$ and let $u = e^{2dz/\sigma^2}$.  Applying Theorem \ref{thm:ddm} we have
\begin{align}
	p& = \frac{1-u^{-1}}{1-u^{-2}}\nonumber\\
	 & =\frac{1}{1+u^{-1}}\nonumber
\end{align}
and
\begin{align*}
	\mu_a & =\frac{2K^2 u^{1/2} }{p\pi^3\sigma^2}\sum_{n=1}^{\infty} \frac{n\sin\paranth{\frac{\pi  n}{2}}}{\paranth{n^2+\paranth{\frac{dK}{\pi\sigma^2}}^2}^2}   \\ 
	  \mu_b & = \frac{2K^2 u^{-1/2}}{(1-p)\pi^3\sigma^2}\sum_{n=1}^{\infty} \frac{n\sin\paranth{\frac{\pi n}{2}}}{\paranth{n^2+\paranth{\frac{dK}{\pi\sigma^2}}^2}^2}.
\end{align*}
  Then to show the equivalence of $\mu_a$ and $\mu_b$, we only need to show that
$u^{1/2}/p = u^{-1/2}/(1-p)$.  Using the above expression for $p$ we obtain
\begin{align*}
	\frac{u^{1/2}}{p} & = u^{1/2}(1+u^{-1})\\
	&= u^{-1/2}+u^{1/2}
\end{align*}
and
\begin{align*}
	\frac{u^{-1/2}}{1-p} & = u^{-1/2}\frac{1+u^{-1}}{u^{-1}}\\
	                   & = u^{-1/2}+u^{1/2}.
\end{align*}

%%%%%%%%%%%%%%%%%%%%%%%%%%%%%%%%%%%%%%%%%%%%%%%%%%%%%%%%%%%%%%%%%%%%%%%%%%%%%%%%
\section{Proof of Theorem \ref{thm:bias}}
The estimate of $w$ is given by $\widehat{w} = (N_1+1)/(N+2)$.  Taking the expectation
of $\widehat{w}$ over $C_u$ and $\epsilon_u$ we have
\begin{align*}
	\mathbf E\bracket{\widehat{w}} & = \frac{\mathbf E\bracket{N_1}}{N+2}+\frac{1}{N+2}\\
																& = \frac{1}{N+2}\sum_{u=1}^N\mathbf E\bracket{\mathbf E\bracket{C_u|\epsilon_u}}+\frac{1}{N+2}\\
	                               & = \frac{1}{N+2}\sum_{u=1}^N\mathbf E\bracket{\frac{w+\epsilon_u}{1+2\epsilon_u}}+\frac{1}{N+2}\\
																 & = \frac{N}{N+2} w\mathbf E\bracket{\frac{1}{1+2\epsilon_u}}+
																     \frac{N}{N+2}\mathbf E\bracket{\frac{\epsilon_u}{1+2\epsilon_u}}+\frac{1}{N+2}\\
																 & = \frac{N}{N+2} w\paranth{1-2\mathbf E\bracket{\frac{\epsilon_u}{1+2\epsilon_u}}}+
																     \frac{N}{N+2}\mathbf E\bracket{\frac{\epsilon_u}{1+2\epsilon_u}}+\frac{1}{N+2}\\	
																&=\frac{N}{N+2}\paranth{w(1-2B)+B}+\frac{1}{N+2}.
\end{align*}
Taking the limit as $N$ goes to infinity, we find that the bias is
\begin{align*}
\lim_{N\rightarrow\infty}	\mathbf E\bracket{\widehat{w}}-w& = \paranth{w(1-2B)+B}-w\\
																													& = B(1-2w).
\end{align*}
%%%%%%%%%%%%%%%%%%%%%%%%%%%%%%%%%%%%%%%%%%%%%%%%%%%%%%%%%%%%%%%%%%%%%%%%%%%%%%%%%%%%%%
\section{Details of Metropolis-Within-Gibbs Sampler}\label{appendix:mcmc}
We use a Metropolis-within-Gibbs scheme to sample from the posterior
distribution of the model parameters. We define the set of model parameters
as $\Theta$ and for any parameter 
 $\kappa\in\Theta$ we define the set of
parameters excluding $\kappa$ as $\Theta_{-\kappa}$. We also define the  observed choice decisions and
response times as $\mathbf C$ and $\mathbf T $. We must sample from the conditional
distribution $\mathbf P(\kappa|\mathbf C, \mathbf T,\Theta_{-\kappa})$
 for each model parameter. We now derive
these conditional distributions and show how to sample from them.

%%%%%%%%%%%%%%%%%%%%%%%%%%%%%%%%%%%%%%%%%%%%%%%%%%%%%%
\subsection{Global Parameters }
The global parameters can be divided into to categories:  means and variances.  We begin the with mean parameters.  For each
mean parameter $\kappa\in\curly{A,\tau,\gamma,\rho,\epsilon}$ we use an uninformative prior distribution that is normal with mean 0 and variance $\sigma^2_{0,\kappa} = 100^2$. The conditional distribution of $\kappa$ is again normal, so it can be directly sampled.  We assume there are $M$ users
and define $\bar{\kappa} = M^{-1}\sum_{u=1}^M\kappa_u$.  The the posterior mean and variance of $\kappa$ are
\begin{align}
	\mu_\kappa' &= \bar{\kappa}\paranth{1+ \frac{\sigma_{\kappa}^{2}}{M\sigma_{0,\kappa}^2}}^{-1}\nonumber\\
	\sigma'^2_{0,\kappa} &= \sigma_{0,\kappa}^2\paranth{M+ \frac{\sigma_{\kappa}^{2}}{\sigma_{0,\kappa}^2}}^{-1}\nonumber
\end{align}
For each variance parameter  $\sigma^2_\kappa\in\curly{\sigma^2_A,\sigma^2_\tau,\sigma^2_\gamma,\sigma^2_\rho,\sigma^2_\epsilon}$ 
the prior distribution of $\sigma_\kappa^2$ is inverse-gamma with shape and scale parameters $a_{\kappa} = 1$ and $b_{\kappa} = 1$, respectively. We can directly sample from the conditional distribution for $\sigma_\kappa^2$ because it is again inverse-gamma with shape parameter $a_{\kappa}'$ and scale parameter $b_{\kappa}'$ given by:
\begin{align}
	a_{\kappa}' &= a_{\kappa} + \frac{M}{2}\nonumber\\
	b_{\kappa}' &= b_{\kappa} + \frac{1}{2} \sum(\kappa_u - \kappa)^2\nonumber.
\end{align}
%%%%%%%%%%%%
\subsection{User Parameters }
We now derive the conditional distribution of the user parameters.
There are two categories of user parameters: those that do not directly affect
the choice decision, ($\curly{A_u,\tau_u,\gamma_u,\rho_u}$) and those
that do ($\epsilon_u$).  We first consider the parameters not involved in the choice
decision.  For each $\kappa_u \in \curly{A_u,\tau_u,\gamma_u,\rho_u}$, 
the posterior is not conjugate so we must use a Metropolis-Hastings step.
The prior distribution of $\kappa_u$ is a normal distribution with mean $\kappa$ and variance $\sigma_\kappa^2$. The conditional distribution of $\kappa_u$ is 
\begin{align}
	\mathbf P\paranth{\kappa_u | \Theta_{-\kappa_u},\mathbf w, \mathbf C, \mathbf T} 
	                       &\propto \mathbf P\paranth{\kappa_u} \mathbf P\paranth{\mathbf T | \Theta,  \mathbf w}\nonumber\\
	                       &\propto e^{-\frac{(\kappa_u - \kappa)^2}{2\sigma_\kappa^2}} 
												          \prod_{s=1}^{S}\prod_{i,j=1}^{N}\prod_{k=1}^{N_{suij}}  g(T^k_{suij}|\tau_{u},\mu_{suij})\nonumber
\end{align}
where $\mu_{suij}$ is given by equation \eqref{eq:musuij}, and $g(t|a,b)$ is the probability density function of a hypoexponential random variable with parameters $a$ and $b$ given by equation \eqref{eq:hypoexponential}.  To sample from this conditional distribution, we use a random walk Metropolis-Hasting step. We define the $l$th sample of $\kappa_u$ as $\kappa_{ul}$, and the proposal for sample $(l+1)$ is drawn from a normal distribution with mean $\kappa_{ul}$ and standard deviation 0.02, where 0.02 is chosen to balance the acceptance rate with step size. 

The conditional distribution of $\epsilon_u$ is
%%%%%
\begin{align}
	\mathbf P\paranth{\epsilon_u | \Theta_{-\epsilon_u},\mathbf w, \mathbf C, \mathbf T} 
	                       &\propto \mathbf P\paranth{\epsilon_u} \mathbf P\paranth{\mathbf C, \mathbf T | \Theta,  \mathbf w}\nonumber\\
	                       &\propto e^{-\frac{(\epsilon_u - \epsilon)^2}{2\sigma_\epsilon^2}} 
												         \prod_{s=1}^{S} \prod_{i,j=1}^{N}\prod_{k=1}^{N_{suij}}  p_{suij}^{C^k_{suij}}\paranth{1-p_{suij}}^{1-C^k_{suij}} g(T^k_{suij}|\tau_{u},\mu_{suij})\nonumber
\end{align}
where $p_{uij}$ is given by equation \eqref{eq:psuij}, $\mu_{suij}$ is given by equation \eqref{eq:musuij}, and $g(t|a,b)$ is the probability density function of a hypoexponential random variable with parameters $a$ and $b$ given by equation \eqref{eq:hypoexponential}.  To sample from this conditional distribution, we use a random walk Metropolis-Hasting step. We define the $l$th sample of $\epsilon_u$ as $\epsilon_{ul}$, and the proposal for sample $(l+1)$ is drawn from a normal distribution with mean $\kappa_{ul}$ and standard deviation 0.02, where 0.02 is chosen to balance the acceptance rate with step size.

%%%%%%%%%%%%%%%%%%%%%%%%%%%%%%%%%%%%%%%%%%%%%%%%%%%%%%%%%%%%%%%%%%%%%%%%%%%%
\subsection{Item Utilities}
For each poll $s$ with $N_s$ items, we use the normalization
$\sum_{i=1}^{N_s}w_{si} = 1$.  Therefore, we only need to sample $w_{si}$ for
$1\leq i \leq N_s-1$ and set $w_{sN_s}=1-\sum_{i=1}^{N_s}w_{si}$.
We use  an uninformative prior distribution for $w_{si}$ that is uniformly distributed between 0 and 1. 
With $M$ users' observed data, the conditional distribution of $w_{si}$ is given by
\begin{align}
	\mathbf P\paranth{w_{si} | \Theta_{-w_{si}}, \mathbf C, \mathbf T} &\propto P\paranth{\mathbf C,\mathbf T | \Theta,  \mathbf w}\nonumber\\
	&\propto \prod_{j=1}^{N}\prod_{u=1}^{M} \prod_{k=1}^{N_{suij}}p_{suij}^{C^k_{suij}}\paranth{1-p_{suij}}^{1-C^k_{suij}}  g(T^k_{suij}|\tau_{u},\mu_{suij})\nonumber
\end{align}
%%%
To sample from this conditional distribution, we use a random walk Metropolis-Hasting step. 
We define the $l$th sample of $w_{si}$ as $w_{sil}$, and the proposal for sample
$(l+1)$  is drawn from a normal distribution with mean $w_{sil}$ and standard deviation 0.02, where 0.02 is chosen to balance the acceptance rate with step size.

\bibliographystyle{ormsv080}
\bibliography{references}
\end{document}